\documentclass[12pt]{article}

\usepackage{graphicx}

\usepackage{color} 
\usepackage{slashed}
\usepackage{amsmath}
\usepackage[pdfstartview=FitH,colorlinks=true,linkcolor=black,anchorcolor=black,citecolor=black,urlcolor=black]{hyperref}
\usepackage{amsmath,amssymb,titling,authblk}
\usepackage{slashed}
\usepackage{amsmath}
\usepackage{amscd}
\usepackage[normalem]{ulem}
\usepackage{appendix}
\usepackage{bbold}

\usepackage[utf8]{inputenc}
\usepackage{slashed}
\usepackage[top=2.3cm,right=2.3cm,left=2.3cm,bottom=2.3cm]{geometry}

\definecolor{verde}{cmyk}{.83,.21,1,.08}
\definecolor{darkorchid}{rgb}{0.6, 0.2, 0.8}
\definecolor{darkgreen}{rgb}{0,.5,0}

\newcommand{\dsf}{{\mathsf{d}}}
\newtheorem{proposition}[equation]{Proposition}

\def\({\left(}
\def\){\right)}
\def\[{\left[}
\def\]{\right]}

\newcommand{\dd}{\mathrm{d}}

\newcommand{\be}{\begin{equation}}
\newcommand{\ee}{\end{equation}}
\newcommand{\bea}{\begin{eqnarray}}
\newcommand{\eea}{\end{eqnarray}}

\newcommand{\la}{\label}

\begin{document}

\title{\Large\bf  On the L$_\infty$ structure of Poisson gauge theory}

\author[1]{O. Abla}
\author[2,3]{V. G. Kupriyanov}
\author[4,5]{M. Kurkov}
\affil[ ]{}
\affil[1]{\textit{\footnotesize CCNH - Universidade Federal do ABC, 09210-580, Santo Andr\'e, SP, 
Brazil. }}
\affil[2]{\textit{\footnotesize CMCC - Universidade Federal do ABC, 09210-580, Santo Andr\'e, SP, 
Brazil. }}
\affil[3]{\textit{\footnotesize Phisics Department, Tomsk State University, 634050, Tomsk, Russia}}
\affil[4]{\textit{\footnotesize INFN-Sezione di Napoli, Complesso Universitario di Monte S. Angelo Edificio 6, via Cintia, 80126 Napoli, Italy.}}
\affil[5]{\textit{\footnotesize Dipartimento di Fisica ``E. Pancini'', Universit\`a di Napoli Federico II, Complesso Universitario di Monte S. Angelo Edificio 6, via Cintia, 80126 Napoli, Italy.}}
\affil[ ]{}
\affil[ ]{\footnotesize e-mail: \texttt{olavoabla1@gmail.com, vladislav.kupriyanov@gmail.com, max.kurkov@gmail.com}}

\maketitle

\begin{abstract}
The Poisson gauge theory is a semi-classical limit of  full non-commutative gauge theory. In this work we construct an L$_\infty^{full}$ algebra which governs both the action of gauge symmetries and the dynamics of the Poisson gauge theory. We derive the minimal set of non-vanishing $\ell$-brackets  and prove that they satisfy the corresponding homotopy relations.  On the one hand, it provides new explicit non-trivial examples of L$_\infty$ algebras.  On the other hand, it can be used as a starting point for bootstrapping the full non-commutative gauge theory. The first few brackets of such a theory are constructed explicitly in the text. In addition we show that the derivation properties of $\ell$-brackets on L$_\infty^{full}$ with respect to the truncated product on the exterior algebra are satisfied only for the canonical non-commutativity. In general, L$_\infty^{full}$ does not have a structure of P$_\infty$ algebra.
\end{abstract}
\section{Introduction}

  {Higher algebraic structures such as  L$_\infty$  and  A$_\infty$ algebras  \cite{Stasheff1,Zwiebach:1992ie,Stasheff2}, exhibit a growing interest, motivated by important applications in physics and mathematics}.
In the physical literature L$_\infty$ algebras appeared for the first time in closed
string field theory \cite{Zwiebach:1992ie} as generalized gauge symmetries. In such theories the commutator of two gauge transformations is again a gauge transformation, however, with a field dependent gauge parameter.
 {This circumstance} weakens the closure constraint, thereby suggesting to generalize the  {very} notion of a Lie algebra.  An appropriate generalization is given by the L$_\infty$ algebras where one has not only a two-bracket (the commutator) but more general multilinear $n$-brackets $\ell_n$ with $n$ inputs. These brackets should satisfy the L$_\infty$ relations also known as higher Jacobi identities. In particular, the usual Jacobi identity
for the two-bracket can be violated by ``total derivative''
terms, thus allowing a mild form of non-associativity.   {The same may happen with the Leibniz rule.} The price to pay is the presence of the higher brackets.

 {The L$_\infty$ algebras do not 
show up in the string field theory only. They also naturally arise  in ``conventional" gauge theories,
like Chern-Simons (CS) and Yang-Mills (YM) theories \cite{Zeitlin:2007vv}, see \cite{Hohm:2017pnh}} for review.
 {In these cases} the structure is  considerably  truncated, and  only a finite number
of higher products and relations  {is} non-trivial.  {At} the classical level L$_\infty$ structure contains all necessary information about the  {gauge} theory, including the gauge symmetry, the field equations and the Noether identities \cite{Stasheff3}. 
According to the conjecture formulated in  \cite{Hohm:2017pnh}, every consistent
gauge theory should be governed by an underlying
L$_\infty$ algebra. 

 {One of the most recent approaches to non-commutative deformations of gauge theories was developed in \cite{BBKL}-\cite{Kup27}. Being based on the use of L$_\infty$ algebras, this formalism allows to deal with a non-constant NC-parameter $\Theta(x)$.} Technically the approach follows the bootstrap logic. {A given undeformed gauge theory defines the initial lower order brackets $\ell_1$ and $\ell_2$.}  For instance, for $U(1)$ gauge theory we set, 
\be\label{i1}
\ell_1(f) :=\dsf f, \quad\mbox{and}\qquad
\ell_1(A) :=\dsf A\,,
\ee
which defines the undeformed gauge variation, $\delta^0_f A=\dsf f$, and the coresponding field strength, $F_0=\dsf A$. The non-commutative deformation is encoded in the theory by identifying the two-bracket  {with a} star commutator of two gauge parameters, 
\be\label{i2}
\ell_2^{NC}(f,g):=i[f,g]_\star,
\ee
for some given star product,
\begin{equation}
f\star g=f\,g+\frac{i\hbar}{2}\,\{f,g\}+{\cal O}\left(\hbar^2\right)\,,\qquad\mbox{with,}\qquad \{f,g\}=\Theta^{ij}(x)\,\partial_if\,\partial_jg\,.
\end{equation}
 {After that, one has to build a
 complete set of $\ell$-brackets of an L$_\infty$ algebra corresponding to the deformed theory.}  {The} L$_\infty$ identities comprising the initially given $\ell$-brackets turn into the equations for definition of the additional brackets,  {like} $\ell_2^{NC}(f,A)$, $\ell_2^{NC}(A,A)$, $\ell_3^{NC}(f,A,A)$  {and} etc. These brackets provide the  {deformed gauge variations}, 
 \be
 \delta_f^{NC}A=\dsf f+\ell_2^{NC}(f,A)+\dots,
 \ee
 closing the NC gauge algebra, 
\be\label{closurefull}
[\delta_f^{NC},\delta_g^{NC}]=\delta_{-i[f,g]_\star}^{NC},
\ee
 and the deformed field strength, 
  \be\label{Ffull}
  {\cal F}_{NC}=\dsf A-\ell_2^{NC}(A,A)/2+\dots, 
  \ee
   {which transforms in a covariant way upon the deformed gauge transformations} 
  \be\label{gccfull}
  \delta_f^{ {NC}}{\cal F}_{NC}=-i[{\cal F}_{NC},f]_\star.
  \ee 
In other words,  {for a given undeformed gauge theory and a given star product},  imposing the guiding principle 
of an underlying L$_\infty$ algebra, we can algebraically bootstrap
all necessary corrections,  {rendering the deformed field strength and the deformed gauge transformations, which close the non-commutative gauge algebra.} We stress that L$_\infty$-formalism is a powerful tool for the construction of perturbative (order by order in $\Theta$) deformations \cite{Kupriyanov:2019ezf,Kup27}. Though, to get an explicit all order expressions normally one needs to invoke additional considerations. 

 {It is a technically complicated task to elaborate the above construction in full generality. For example, the expression for the star product and star commutator is very complex and is known in a precise form only in a few number of examples of non-constant $\Theta$. Therefore from now on we adopt some simplifications.} Instead of dealing with the full non-commutative gauge algebra one may work with its semi-classical limit, which is given by the Poisson gauge algebra \cite{KS21}. In this algebra the commutator of two gauge transformations with gauge parameters $f$ and $g$ is a gauge transformation corresponding to the Poisson bracket of these parameters\footnote{Remind that the Poisson bracket is the semi-classical limit of the star commutator, $\{f,g\}=\lim_{\hbar\to0}[f,g]_\star/i\hbar$.}, 
\begin{equation}
[\delta_f,\delta_g]A=\delta_{\{f,g\}}A\,.\label{ga}
\end{equation}
 {A dynamical field theoretical model having the Poisson gauge algebra \eqref{ga} as a corresponding algebra  of gauge symmetries is called the Poisson gauge theory \cite{Kup33}. This model is designed to investigate the semi-classical features of the full non-commutative gauge theory with coordinate dependent non-commutativity $\Theta^{ij}(x)$. }

  {It is remarkable, that there is another way  to build the Poisson gauge theory, which does not rely explicitly on the L$_\infty$-structures. In particular,}
 in \cite{KS21} it was proposed an approach to the construction of (almost)-Poisson gauge transformations and the corresponding gauge algebra, based on the symplectic embeddings of (almost)-Poisson structures. The further generalization of the symplectic embedding formalism to the construction of the Poisson gauge theory was elaborated in \cite{Kup33}.
The two approaches for non-commutative gauge theories, mentioned above, can be seen as complementary. As it has already been mentioned the homotopy algebras provide the perturbative in $\Theta$ corrections. Once we know the structure of these corrections we may employ the symplectic embedding framework which is especially good for obtaining explicit all-order in $\Theta$ form of the deformed constructions.  

The aim of the  {present} work is to esteblish the  {direct connection between the two approaches to Poisson gauge theories: the} formulation in terms of L$_\infty$-algebras \cite{BBKL}-\cite{Kup27} and the symplectic embedding formalism \cite{KS21,Kup33}. More precisely, we will construct the L$_\infty^{full}$ algebra corresponding to  {a generic} Poisson gauge theory \cite{Kup33}. A part of this work has  {already been} done in \cite{KS21}, where the algebra L$_\infty^{gauge}$ describing the action of Poisson gauge symmetries on the gauge field, was constructed. The new results of the  {present} research include the description of a dynamical sector of the Poisson gauge theory in terms of L$_\infty$ algebras and investigation of the corresponding P$_\infty$ structure.

The paper is organized as follows. In Sec. 2 we briefly review the construction of the Poisson gauge theory in the symplectic embedding formalism. We start Sec. 3 providing the basic definitions of L$_\infty$ algebra.  After that, in Proposition \ref{pr2}, we represent the L$_\infty$-relations in the form, which is convenient for our purposes. The main result is formulated and proven in  Proposition \ref{pr3}. The explicit examples of $\mathfrak{su}(2)$-like structure and $\kappa$-Minkowski deformations are analyzed Sec. 4. In Sec. 5 we discuss the corresponding P$_\infty$ structure. The last section is dedicated to the explicit construction of the lower order $\ell$-brackets corresponding to the full non-commutative gauge theory. 

\section{Poisson gauge theory}

The aim of this section is to review the construction of the field theoretical model which has a Poisson gauge algebra (\ref{ga}) as a corresponding algebra of gauge symmetries. We start with the definition of the Poisson gauge transformations $\delta_fA$ which should satisfy the following two conditions: the closure of the algebra (\ref{ga})
and the correct commutative limit to the standard $U(1)$ gauge transformations, $\lim_{\Theta\to0}\delta_fA=\partial f$. 
If $\Theta^{ij}$ is constant, one may easily see that the straightforward expression, $\delta^{can}_fA=\partial f+\{A,f\}$, satisfies (\ref{ga}). However, for non-constant $\Theta^{ij}(x)$ the standard Leibniz rule with respect to the partial derivative is violated, $\partial\{f,g\}\neq\{\partial f,g\}+\{f,\partial g\}$, therefore the same expression will not close the algebra (\ref{ga}) anymore. To overcome this problem one has to modify the expression for the gauge transformations introducing the corrections proportional to the derivatives of the non-commutativity $\partial \Theta^{ij}(x)$ which would compensate the violation of the Leibniz rule. 

 The problem with the violation of the Leibniz rule can be solved in an extended space \cite{KS21}. To each coordinate $x^i$ we introduce a conjugate variable $p_i$, in such a way that the corresponding Poisson brackets,
\begin{equation}\label{PB1}
\{x^i,x^j\}=\Theta^{ij}(x)\,,\qquad\{x^i,p_j\}=\gamma^i_j(x,p)\,,\qquad \{p_i,p_j\}=0\,,
\end{equation}
should satisfy the Jacobi identity. For constant $\Theta^{ij}$ one finds, $\gamma^i_j(x,p)=\delta^i_j$, so the Poisson bracket, $\{f(x),p_i\}=\partial_if(x)$, is just a partial derivative of this function. In case if $\Theta^{ij}(x)$ is  not constant the expression for $\gamma^i_j(x,p)$ is more complicated. The Jacobi identity for the algebra (\ref{PB1}) implies the partial differential equation on the matrix $\gamma^i_j(x,p)$,
\begin{equation}
\gamma^l_b\,\partial^b_p\gamma_a^k-\gamma^k_b\,\partial^b_p\gamma^l_a+\Theta^{lm}\,\partial_m\gamma_a^k-\Theta^{km}\,\partial_m\gamma_a^l-\gamma^m_a\,\partial_m\Theta^{lk}=0\,,\label{first}
\end{equation}
with $\partial_m=\partial/\partial x^m$ and $\partial^b_p=\partial/\partial p_b$. Perturbative solution is given by,
\begin{eqnarray}
\gamma^k_a(x,p)&=&\sum_{n=0}^\infty  \gamma_{ a }^{ k (n)} \label{gps}\\
&=&\delta^k_a-\frac{1}{2}\, \partial_a \Theta^{kb} p_b-\frac{1}{12}\left(2\,\Theta^{cm}\partial_a\partial_m\Theta^{bk}+\partial_a\Theta^{bm}\partial_m\Theta^{kc}\right)p_bp_c+{\cal O}(\Theta^3)\,.\notag
\end{eqnarray}
The recursive formula for the construction of the functions, $\gamma_{ a }^{ k (n)} = \gamma_{ a }^{ k| j_1\dots j_n }(x) \,p_{j_1}\dots p_{j_n}$, can be found in \cite{Kup21}. 

Instead of working with standard partial derivatives $\partial_i$ in the definition of the gauge transformation one may introduce the `twisted' derivative of function $f(x)$ defined in terms of Poisson bracket with $p$-variable, $\partial^\tau_if(x):=\{f(x),p_i\}=\gamma^l_i(x,p)\partial_lf(x)$. The Jacobi identity for the Poisson brackets (\ref{PB1}) and the fact that $\{p_i,p_j\}=0$ imply two important properties of `twisted' derivatives: they commute, $\partial^\tau_i\partial^\tau_j=\partial^\tau_j\partial^\tau_i$, and satisfy the Leibniz rule, $\partial^\tau_a\{f,g\}=\{\partial^\tau_af,g\}+\{f,\partial^\tau_ag\}$.
The price to pay however is that the expression $\{f(x),p_i\}$ depends also on the auxiliary non-physical $p$-variables. It turns out that the auxiliary variables can be eliminated in the consistent way by introducing the constraints, $p_a=A_a(x)$. In \cite{KS21} it was demonstrated that the gauge transformations given by,
\begin{equation}\label{gtA}
\delta_f A_a=\gamma^l_a(A)\,\partial_lf(x)+\{A_a(x),f(x)\}\,,
\end{equation}
where, $\gamma^l_a(A):=\gamma^l_a(x,p)|_{p_a=A_a(x)}$, close the algebra (\ref{ga}) and have the desired commutative limit.

To be consistent with the semi-classical limit the gauge variation of the matter field $\psi$ is defined as, $\delta_f\psi=\{\psi,f\}$. The gauge covariant derivative of matter field is the object which should satisfy the following two conditions: it should transform covariantly under the gauge transformation, 
\begin{eqnarray}
\delta_f{\cal D}_a(\psi)=\{{\cal D}_a(\psi),f\}\,,\label{w6a}
\end{eqnarray}
 and reproduce the standard derivative in the commutative limit, $\lim_{\Theta\to0}{\cal D}_a(\psi)=\partial_a\psi\,.$ Such an object can be constructed following \cite{Kup33}:
\begin{equation}\label{w6}
{\cal D}_a(\psi)=\rho_a^i(A)\left(\gamma_i^l(A)\,\partial_l\psi+\{A_a,\psi\}\right)\,,
\end{equation}
where the matrix, $\rho_a^i(A):=\rho^i_a(x,p)|_{p_a=A_a(x)}$, with,
\begin{eqnarray}
\gamma^j_b\,\partial^b_p\, \rho_a^i+\rho_a^b\,\partial^i_p\gamma^j_b+\Theta^{jb}\,\partial_b\rho_a^i=0\,.\label{second}
\end{eqnarray}
Perturbative solution of this equation reads,
\begin{eqnarray}
\rho^i_a(x,p)&=&\sum_{n=0}^\infty\rho_{ a }^{ i (n)}\\
&=&\delta^i_a-\frac{1}{2}\, \partial_a \Theta^{ib} p_b+\frac{1}{6}\left(2\,\Theta^{cm}\partial_a\partial_m\Theta^{ib}-\partial_a\Theta^{bm}\partial_m\Theta^{ic}\right)p_bp_c+{\cal O}(\Theta^3)\,.\notag
\end{eqnarray}
For constant $\Theta^{ij}$ one finds just, $\rho^i_a(x,p)=\delta^i_a$.

The commutator relation for the covariant derivatives,
\begin{eqnarray}
\left[{\cal D}_a,{\cal D}_b\right]=\{{\cal F}_{ab},\,\cdot\,\}+\left({\cal F}_{ad}\,{\Lambda}_b{}^{de}-{\cal F}_{bd}\,{\Lambda}_a{}^{de}\right){\cal D}_e\,,\label{comm}
\end{eqnarray}
where, ${\Lambda}_b{}^{de}=\left(\rho^{-1}\right)_j^d\left(\partial^j_A\rho_b^m-\partial^m_A\rho_b^j\right)\left(\rho^{-1}\right)_m^e\,,$ defines the Poisson field strength:
\begin{eqnarray}
{\cal F}_{ab}:=\frac12\left(\rho^c_a\rho^d_b-\rho^c_b\rho^d_a\right)\left(\gamma_c^m\,\partial_m A_d+\{A_c,A_d\}\right)\,,\label{pfs}
\end{eqnarray}
which  transforms covariantly, \begin{equation}
\delta_f {\cal F}_{ab}=\{{\cal F}_{ab},f\}\,.\label{gcc}
\end{equation} and reproduces the abelian field strength in the commutative limit,
\begin{equation}
{\lim_{\Theta\to0}\,{\cal F}_{ab} =\partial_a A_b-\partial_bA_a\,.}
\end{equation}

Now we turn to the consistent gauge covariant deformation of the equations of motion.Taking into account (\ref{gcc}) and (\ref{w6a}) the straightforward candidate reads,
\begin{equation}\label{Es}
{\cal E}_{S}^b:={\cal D}_a\left({\cal F}^{ab}\right)=0\,,
\end{equation}
where the subscript $S$ stands for ``straightforward''. This quantity transforms covariantly, $\delta_f{\cal E}_{S}^b=\{{\cal E}_{S}^b,f\}$, and reproduces the first pare of Maxwell equations in the commutative limit, $\lim_{\Theta\to0}{\cal E}_{S}^b=\partial_aF^{ab}_0$. These two are essential requirements that we should impose on the equations of motion of the deformed theory. However, they are not the only existing. When we start working with the equations of motion it is not always guaranteed that these equations can be obtained from the action principle. 

The alternative way of obtaining of the deformed equations of motion from the corresponding deformed Lagrangian was proposed in \cite{Kupriyanov:2020sgx}. Basically the idea is the following, having in hands the gauge covariant Poisson field strength (\ref{pfs}) and the covariant derivative (\ref{w6}) one constructs the gauge covariant deformation of the standard Lagrangian,
\begin{equation}
{\cal L}_{g}=-\frac14\,{\cal F}_{ab}\,{\cal F}^{ab}\,,\qquad\mbox{with,}\qquad \delta_f{\cal L}_{g}=\{{\cal L}_{g},f\}\,.
\end{equation}
Then introducing an appropriate measure $\mu(x)$, such that for any two Schwartz functions $f$ and $g$ holds,
\begin{equation}\label{measure}
\int\!d^Nx\,\mu(x)\,\{f,g\}=0\,\qquad  \Leftrightarrow \qquad \partial_l\left(\mu(x)\,\Theta^{lk}(x)\right)=0\,,
\end{equation}
one constructs the gauge invariant action,
\begin{equation}\label{ag}
S_g=\int\!d^Nx\,\mu(x)\,{\cal L}_{g}\,,\qquad \mbox{such that,}\qquad \delta_f S_g\equiv0\,.
\end{equation}
Note that in general the measure function $\mu(x)$ is not constant, moreover in $\kappa$-Minkowski case, see e.g. \cite{Jonke1,KKV}, it does not tend to the trivial measure in the commutative limit. Therefore the action (\ref{ag}) does not have a desired commutative limit, $\lim_{\Theta\to0}\mu(x)\,{\cal L}\neq{\cal L}^0$. This issue was addressed in \cite{Jonke1} by introducing the gauge covariant factor in the Lagrangian density which would ``absorb" the non-trivial measure. Another possibility is to rescale the original non-commutativity in such a way that the measure becomes trivial, see \cite{KKV} for more details. 

 The Euler-Lagrange equations following from the action (\ref{ag}),
\begin{equation}\label{EL1}
{\cal E}^b_{EL}:=\frac{\delta S_g}{\delta A_b}=0\,,
\end{equation}
by construction are gauge covariant, in a sense that the solutions are mapped onto the solutions under the gauge transformations. The subscript $EL$ stands for the ``Euler-Lagrange''. Without going into the tedious calculations which can be found in \cite{Kup33} we write here,
\begin{eqnarray}\label{EL}
{\cal E}^b_{EL}=\rho_c^b\left[\mu\,{\cal D}_a\left({\cal F}^{ac}\right)+\frac{\mu}{2}\,{\cal F}^{cb}\,\Lambda_b{}^{de}\,{\cal F}_{de}-\mu\,{\cal F}^{db}\,\Lambda_b{}^{ce}\,{\cal F}_{de}+\left(\rho^{-1}\right)^c_k{\cal F}^{ab}\partial_i\left(\mu\,\rho_a^l\,\rho_b^k\,\gamma^i_l\right)\right]\,,
\end{eqnarray}
where, 
\begin{eqnarray}
\Lambda_b{}^{de}(A)=\left(\rho^{-1}\right)_j^d\left(\partial^j_A\rho_b^m(A)-\partial^m_A\rho_b^j(A)\right)\left(\rho^{-1}\right)_m^e\,.\label{Lambda}
\end{eqnarray}
We stress that the whole theory is completely determined by the two matrices, the matrix $\gamma^l_a(A)$ which should obey by the equation (\ref{first}) and the matrix $\rho_a^i(A)$ satisfying the equation (\ref{second}). 

It is worth mentioning here that for some specific choices of the non-commutativity parameter the equations of motion constructed according to (\ref{Es}) and (\ref{EL}) are equivalent. In particular, for the $su(2)$-like Poisson structure, $\Theta^{ab}(x)=2\,\alpha\,\varepsilon^{ab}{}_c\,x^c$, the integration measure is constant $\mu(x)=1$ and the additional terms in (\ref{EL}) vanish in such a way that,\footnote{In this particular situation the matrix $\rho_a^b(A)$ plays the role of the Lagrangian multiplier in the sense of non-Lagrangian systems \cite{GK}. The original equations of motion, ${\cal E}^a_{S}=0$, are non-Lagrangian, however the multiplication by the non-degenerate matrix $\rho_a^b(A)$ transforms them into a equivalent set of Euler-Lagrange equations (\ref{EL1}) for the action (\ref{ag}). }
\begin{equation}\label{ELS}
{\cal E}^b_{EL}=\rho_a^b\,{\cal E}^a_{S}\,.
\end{equation}
On the other hand, for the $\kappa$-Minkowski non-commutativity the integration measure $\mu(x)$ is non-trivial \cite{KKV}, so the relation (\ref{ELS}) does not hold. In the Section 4 we will discuss these two examples in more details.

\section{Relation to L$_\infty$ algebras}

In this section we discuss the relation of the described above field theoretical model with the L$_\infty$ algebras. We will construct an explicit form of the L$_\infty^{full}$ algebra which governs both kinematical and dynamical sectors of the theory. 

\subsection{Basic definitions and identities} 

The definition of the L$_\infty$ algebra can be resumed as follows. An L$_\infty$-algebra is a $ \mathbb{Z}$-graded vector space $V=\bigoplus_{k\in  \mathbb{Z}}\, V_{k}$ together with a sequence of graded antisymmetric multilinear maps,
\begin{align*}
\ell_n:V^{\otimes n}\to V \ , \quad  v_1\wedge \dots\wedge v_n \longmapsto \ell_n (v_1,\dots,v_n)\,,
\end{align*}
called $\ell$-brackets. The property of graded antisymmetry menas
\begin{equation}\label{eq:gradedantisym}
\ell_n (\dots, v,v',\dots) = -(-1)^{|v|\,|v'|}\, \ell_n (\dots, v',v,\dots) \ ,
\end{equation}
where we denote the degree of a homogeneous element $v\in V$ by $|v|$.
The $\ell$-bracket is a map of degree $|\ell_n|=2-n$, implying that
\begin{equation*}
\big|\ell_n(v_{1}, \dots ,v_{n})\big| = 2-n +\sum_{j=1}^n \, |v_j| \ .
\end{equation*}
The brackets $\ell_n$ should satisfy an (infinite) tower of
identities ${\cal J}_n(v_1,\dots,v_n)=0$ for each $n\geq1$, called homotopy relations, with
\begin{eqnarray} \label{eq:calJndef}
{\cal J}_n(v_1,\dots,v_n) &:=& \sum^n_{i=1}\, (-1)^{i\,(n-i)} \
  \sum_{\sigma\in{\rm Sh}_{i,n-i}} \,  \chi(\sigma;v_1,\dots,v_n) \\ &&
                                                                      \qquad
                                                                      \times
                                                                      \ell_{n+1-i}\big(
                                                                      \ell_i(v_{\sigma(1)},\dots,
                                                                      v_{\sigma(i)}),
                                                                      v_{\sigma(i+1)},\dots
                                                                      ,v_{\sigma(n)}\big)
                                                                      \
                                                                        , \notag
\end{eqnarray}
where, for each $i=1,\dots,n$, the second sum runs over $(i,n-i)$-shuffled permutations
$\sigma\in S_n$ of degree $n$ which are restricted as 
\begin{equation*}
\sigma(1)<\dots<\sigma(i) \qquad \mbox{and} \qquad \sigma(i+1)<\dots < \sigma(n) \ .
\end{equation*}
The Koszul sign $\chi(\sigma;v_1,\dots,v_n)=\pm\, 1$ is determined from the grading by
\begin{align*}
v_{\sigma(1)}\wedge\cdots\wedge v_{\sigma(n)} = \chi(\sigma;v_1,\dots,v_n) \ v_1\wedge\cdots\wedge v_n \ .
\end{align*}

In particular, the first relation ${\cal J}_1(v):=\ell_1(\ell_1(v))=0$ signifies that $\ell_1$ is a
differential making $(V,\ell_1)$ into a cochain complex, while the second identity,
\begin{equation*}
{\cal J}_2(v_1,v_2):  = \ell_1\big(\ell_2(v_1,v_2)\big) - \ell_2\big(\ell_1(v_1),v_2\big) - (-1)^{|v_1|}\, \ell_2\big(v_1, \ell_1(v_2)\big)\,,
\end{equation*}
 implies that $\ell_1$ is a (graded)
derivation of the bracket $\ell_2$, i.e. $\ell_2$ is a cochain
map. The identity with three entries schematically written as ${\cal J}_3:=\ell_1\ell_3+\ell_2\ell_2+\ell_3\ell_1=0$ means that the bracket $\ell_2$ obeys the Jacobi identity
up to exact in $\ell_1$ terms, i.e. $\ell_2$ induces a (graded) Lie bracket on the
cohomology of $\ell_1$.

In this work we are dealing with the structure of a
three-term L$_\infty$-algebra describing
the cochain complex,
\begin{eqnarray}\label{eq:cochaincomplex}
V_0\xrightarrow{ \ \ \ell_1 \ \ }V_1\xrightarrow{ \ \ \ell_1 \ \ 
  }V_2
\end{eqnarray}
corresponding to the underlying graded vector space
$$
V=V_{0}\oplus V_{1}\oplus V_{2}  \ .
$$
Physically, $V_0$ corresponds to the space of gauge parameters or the zero-forms $f$. The subspace $V_{1}=\{A\}\cup\{\varphi\}$ contains the gauge fields represented by the $1$-forms, $A=A_a\dsf x^a$, and the matter fields $\varphi$ which are $0$-forms. And finally, 
$V_2=\{E^A\}\cup\{E^\varphi\}\cup\{{\cal E}_{EL}\}$ is the space of covariant objects. The subspace of $2$-forms $\{E^A\}$ contains the field strength, ${\cal F}(A)={\cal F}_{ab}\,\dsf x^a\wedge \dsf x^b$, while the subspace of $1$-forms $\{E^\varphi\}$ contains covariant derivatives of the matter fields, ${\cal D}(\varphi)={\cal D}(\varphi)_a\,\dsf x^a$ and $\{{\cal E}_{EL}\}$ is the subspace of the equations of motion. 

 In what follows we will need the explicit form of some identities which is given by the following Proposition. 
 
 {\proposition \label{pr2}
The L$_\infty$ relations, ${\cal J}_n(f,g,A^{\otimes n-2})=0$,  ${\cal J}_n(f,A^{\otimes n-1})=0$, and ${\cal J}_n(f,\varphi,A^{\otimes n-2})=0$ are given explicitly by,
\begin{eqnarray}\label{Jnfg}
{\cal J}_n(f,g,A^{\otimes n-2})&=&\ell_1\left(\ell_n(f,g,A^{\otimes n-2})\right)-(-1)^{n}\ell_n(\ell_1(f),g,A^{\otimes n-2})\\
&&-(-1)^{n}\ell_n(f,\ell_1(g),A^{\otimes n-2})-(-1)^{n}(n-2)\,\ell_n\left(f,g,\ell_1(A),A^{\otimes n-3}\right)\notag\\
&&+\sum_{i=2}^{n-1}(-1)^{i(n-i)}\left[\binom{n-2}{i-2}\,\ell_{n-i+1}\left(\ell_i(f,g,A^{\otimes i-2}),A^{\otimes n-i}\right)\right. \notag\\
&&-(-1)^{i}\binom{n-2}{i-1}\,\ell_{n-i+1}\left(\ell_i(f,A^{\otimes i-1}),g,A^{\otimes n-i-1}\right)\notag\\
&&+\left.(-1)^{i}\binom{n-2}{i-1}\,\ell_{n-i+1}\left(\ell_i(g,A^{\otimes i-1}),f,A^{\otimes n-i-1}\right) \right] \,,\notag\\
{\cal J}_n(f,A^{\otimes n-1})&=&\ell_1\left(\ell_n(f,A^{\otimes n-1})\right)-(-1)^{n}\ell_n(\ell_1(f),A^{\otimes n-1}) \label{JnfA}\\
&&-(-1)^{n}(n-1)\,\ell_n(f,\ell_1(A),A^{\otimes n-1})\notag\\
&&+\sum_{i=2}^{n-1}(-1)^{i(n-i)}\left[\binom{n-1}{i-1}\,\ell_{n-i+1}\left(\ell_i(f,A^{\otimes i-1}),A^{\otimes n-i}\right)\right. \notag\\
&&+\left.(-1)^{i}\binom{n-1}{i}\,\ell_{n-i+1}\left(\ell_i(A^{\otimes i}),f,A^{\otimes n-i-1}\right)\right]\,, \notag
\end{eqnarray}
and
\begin{eqnarray}\label{Jnfphi}
{\cal J}_n(f,\varphi,A^{\otimes n-2})&=&\ell_1\left(\ell_n(f,\varphi,A^{\otimes n-2})\right)-(-1)^{n}\ell_n(\ell_1(f),\varphi,A^{\otimes n-2})\\
&&-(-1)^{n}\ell_n(f,\ell_1(\varphi),A^{\otimes n-2})-(-1)^{n}(n-2)\,\ell_n\left(f,\varphi,\ell_1(A),A^{\otimes n-3}\right)\notag\\
&&+\sum_{i=2}^{n-1}(-1)^{i(n-i)}\left[\binom{n-2}{i-2}\,\ell_{n-i+1}\left(\ell_i(f,\varphi,A^{\otimes i-2}),A^{\otimes n-i}\right)\right. \notag\\
&&+\binom{n-2}{i-1}\,\ell_{n-i+1}\left(\ell_i(f,A^{\otimes i-1}),\varphi,A^{\otimes n-i-1}\right)\notag\\
&&+\left.(-1)^{i}\binom{n-2}{i-1}\,\ell_{n-i+1}\left(\ell_i(\varphi,A^{\otimes i-1}),f,A^{\otimes n-i-1}\right)\right] \,.\notag
\end{eqnarray}

}

{\it Proof.} Let us first discuss the relations with two gauge parameters $f$ and $g$ and $(n-2)$ gauge fields $A$. The sum in (\ref{eq:calJndef}) is taken over inequivalent splittings of elements. For fixed $n$ and $i$ there are $\binom{n}{i}=n!/(i!(n-i)!)$ splittings of elements, so the sum $\sum_{\sigma\in{\rm Sh}_{i,n-i}}$ contains $\binom{n}{i}$ elements, see \cite{Hohm:2017pnh} for more details. If $i=n$ the contribution to  (\ref{eq:calJndef}) is $\ell_1(\ell_n(f,g,A^{\otimes n-2}))$,  and when $i=1$ the conributions are $\ell_n(\ell_1(f),g,A^{\otimes n-2})$, $\ell_n(f,\ell_1(g)A^{\otimes n-2})$ and also $n-2$ terms of the type $\ell_n\left(f,g,\ell_1(A),A^{\otimes n-3}\right)$. 
In the case $i\neq n$ and $i\neq 1$, taking into account that we have $n-2$ identical entries, the sum $\sum_{\sigma\in{\rm Sh}_{i,n-i}}$  will contain $\binom{n-2}{i-2}$ contributions of the type $\ell_{n-i+1}(\ell_i(f,g,A^{\otimes i-2}),A^{\otimes n-i})$, $\binom{n-2}{i-1}$ contributions of the type $\ell_{n-i+1}\left(\ell_i(f,A^{\otimes i-1}),g,A^{\otimes n-i-1}\right)$, the same number of the terms $\ell_{n-i+1}\left(\ell_i(g,A^{\otimes i-1}),f,A^{\otimes n-i-1}\right)$, 
and also $\binom{n-2}{i}$ elements like $\ell_{n-i+1}(\ell_i(A^{\otimes i}),f,g,A^{\otimes n-i-2})$. The latter however can be set to zero since we are working with the off-shell closure condition for the gauge algebra, see the comment after the eq. (\ref{closure}). All together,
\begin{equation*}
\binom{n-2}{i-2}+\binom{n-2}{i-1}+\binom{n-2}{i-1}+\binom{n-2}{i}=\binom{n}{i}\,,
\end{equation*}
which is exactly what we need. The sign $(-1)^{i-1}$ in the fourth and the fifth lines appeared because the element $g$ which originally was staying in the second slot was moved to the  $(i+1)$-th slot. 

The expression for the identities with one gauge parameter ${\cal J}_n(f,A^{\otimes n-1})$ can be obtained in the same manner. For $i=n$  the contribution to  the sum (\ref{eq:calJndef}) is $\ell_1(\ell_n(f,g,A^{\otimes n-2}))$, when $i=1$ the corresponding terms are $\ell_n(\ell_1(f),A^{\otimes n-1})$ and $(n-1)\ell_n(f,\ell_1(A),A^{\otimes n-2})$ multiplied by the sign factor. In case if $i\neq n$ and $i\neq 1$, the fact that there are $n-1$ identical elements, implies that the sum $\sum_{\sigma\in{\rm Sh}_{i,n-i}}$  will contain $\binom{n-1}{i-1}$ contributions of the type $\ell_{n-i+1}(\ell_i(f,A^{\otimes i-1}),A^{\otimes n-i})$
and also $\binom{n-1}{i}$ elements $\ell_{n-i+1}(\ell_i(A^{\otimes i}),f,A^{\otimes n-i-1})$. All together it gives,
$
\binom{n-1}{i-1}+\binom{n-1}{i}=\binom{n}{i}\,,
$
as expected.  The sign $(-1)^{i}$ in the last line appeared because the element $f$ which originally was staying in the first slot appears now in the $(i+1)$-th slot. 

The derivation of the expression for ${\cal J}_n(f,\varphi,A^{\otimes n-2})$ is absolutely analogous to ${\cal J}_n(f,g,A^{\otimes n-2})$ with only difference that the element $\varphi$ has degree one and therefore its permutations with the elements $A$ do not create a sign factor. $\Box$

\subsection{L$_\infty$ description of Poisson gauge theory}

Now we may proceed to the definition of the corresponding $\ell$-brackets comparing the constructions of the Poisson gauge theory with the corresponding expressions in the L$_\infty$-formalism. We start with the gauge variations of the gauge fields $A$ which are defined in the L$_\infty$ framework as,
\begin{eqnarray}
\label{var}
  \delta_{f}  A &:=&\sum^\infty_{n= 0}   {1\over n!}
      (-1)^{n(n-1)\over 2}\,
 \ell_{n+1}(f, A, \dots, A )\notag\\&=&\dsf f+\ell_2(f,A)-\frac12\ell_3(f,A,A)+\dots\, .
\end{eqnarray}
The homotopy relations with two gauge parameters, ${\cal J}_{n+2}(f,g,A^{\otimes n})=0$, imply the gauge closure condition \cite{Stasheff2,Hohm:2017pnh}, 
\begin{eqnarray}\label{closure}
[\delta_{f},\delta_{g}]A
  = \delta_{[\![f,g]\!]_A} A +\delta^T_{C(f,g,E,A)}A\ ,
  \end{eqnarray}
  where,
  \begin{eqnarray}   
  [\![f,g]\!]_A: = -\sum_{n=0}^\infty \, \frac1{n!}\,
 (-1)^{n(n-1)\over 2}\, \ell_{n+2}\left(f,g,A^{\otimes n} \right)=-\ell_2(f,g)-\ell_3(f,g,A)+\dots\,,\notag
\end{eqnarray}
and the object $C(f,g,E,A)$ is constructed from the brackets of the type $\ell_{n+3}(f,g,E,A^{\otimes n})$ with $E=0$ being the equations of motion. The term $\delta^T_{C(f,g,E,A)}A$ accounts for the on-shell gauge closure. In our case we have the off-shell closure, so we just set, $\ell_{n+3}(f,g,E,A^{\otimes n})=0$, as it was already mentioned in the proof of Proposition \ref{pr2}.
The homotopy relations with three gauge parameters, ${\cal J}_{n+3}(f,g,h,A^{\otimes n})=0$, imply the Jacobi identity for the corresponding gauge algebra, $[\delta_h,[\delta_{f},\delta_{g}]]+\mbox{cycl.}(f,g,h)=0$. 

The Poisson gauge transformations closing tha algebra (\ref{ga}) are determined according to (\ref{gtA}), 
\begin{equation}
\label{h2}
\delta_f A_a= \gamma^k_a(A)\,\partial_kf+\{A_a, f\}\,,
\end{equation}
where the matrix
\be\label{h3}
\gamma_{ a }^{ k }(A) =\sum_{n=0}^\infty \gamma_{ a }^{ k (n)}\,,\qquad \gamma_{ a }^{ k (0)}= \delta_{ a }^{ k }\,,\qquad \gamma_{ a }^{ k (n)} = \gamma_{ a }^{ k| j_1\dots j_n }(x) \,A_{j_1}\dots A_{j_n}\,,
\ee 
should satisfy the equation (\ref{first}).  Comparing (\ref{var}) to (\ref{h2}) and (\ref{closure}) to (\ref{ga}) we conclude that the only non-vanishing brackets coming from this sector read \cite{KS21},
\begin{eqnarray}\label{h6}
\ell_1(f)&=&\dsf f\,,\qquad \ell_2(f,g)=-\{f,g\}\,,\\
 \ell_2(f,A)&=&\left(\gamma^{k(1)}_i\,\partial_kf+\{A_i,f\}\right)\dsf x^i\,,\notag\\
\ell_m(f,A^{\otimes m-1})&=&(m-1)! (-1)^{\frac{(m-1)\,(m-2)}2}\,\gamma_i^{k(m-1)}\partial_kf\,\dsf x^i\,,\qquad m\geq3\,.\notag
\end{eqnarray}
Since the tensor $\gamma_{ a }^{ k| j_1\dots j_n }(x) $ by construction is symmetric in the indices $j_1,\dots,j_n$ the off-diagonal elements read,
\begin{equation}\label{h6off}
\ell_m(f,A^{(1)},\dots,A^{(m-1)})=(m-1)! (-1)^{\frac{(m-1)\,(m-2)}2}\,\gamma_{ i }^{ k| j_1\dots j_{m-1} }A^{(1)}_{j_1}\dots A^{(m-1)}_{j_{m-1}}\partial_kf\,\dsf x^i\,.
\end{equation}
To prove that these brackets indeed determine an L$_\infty$ algebra we have to show that the corresponding homotopy relations ${\cal J}_{n+2}(f,g,A^{\otimes n})=0$, and ${\cal J}_{n+3}(f,g,h,A^{\otimes n})=0$, hold true. Since the only non-vanishing bracket with two gauge parameters is, $\ell_2(f,g)=-\{f,g\}$, the only non-trivial L$_\infty$ relation with three gauge parameters to be checked reads,
\begin{equation*}
{\cal J}_{3}(f,g,h):=\ell_2(\ell_2(f,g),h)+\ell_2(\ell_2(g,h),f)+\ell_2(\ell_2(h,f),g)=0\,.
\end{equation*}
It holds true as a consequence of the Jacobi identity for the Poisson bracket. The situation with the brackets $\ell_m(f,A^{\otimes m-1})$ and the identities with two gauge parameters is more complicated, we will prove it carefully in the Proposition \ref{pr3}.

The gauge variation of the matter field $\varphi$ can be encoded in terms of the brackets $ \ell_{n+2}(f,\varphi, A^{\otimes n} )$ as,
\begin{eqnarray}
\label{varphi1}
  \delta_{f}  \varphi&=&\sum_{n= 0}^\infty   {1\over n!}
      (-1)^{n(n+1)\over 2}\,
 \ell_{n+2}(f,\varphi, A, \dots, A )\\&=&\ell_2(f,\phi)-\ell_3(f,\varphi,A)-{1\over2}\ell_4(f,\varphi,A,A)+\dots\, .\notag
\end{eqnarray} 
Then the homotopy relations, ${\cal J}_{n+3}(f,g,\varphi,A^{\otimes n})=0$, imply the gauge closure condition, 
\begin{equation*}
[\delta_{f},\delta_{g}]\varphi
  = \delta_{[\![f,g]\!]_A}\varphi\,.
\end{equation*}
 In our case, $\delta_f\varphi=\{\varphi,f\}\,.$ So, the additional non-vanishing bracket is, $\ell_2(f,\varphi)=-\{f,\varphi\}$. Again the only non-trivial L$_\infty$ relation with two gauge parameters and one matter field, ${\cal J}_{3}(f,g,\varphi)=0$, is satisfied because of the Jacobi identity for the Poisson bracket.
  
  From now on we take into account the brackets which belong to the subspace of the covariant objects $V_2$. If we define the field strength as the following combination of $\ell$-brackets,
  \begin{eqnarray}\label{FA1}
{\cal F}(A)=\sum^\infty_{n=1} \;{1\over n!}
(-1)^\frac{n(n-1)}{2}\, \ell_n(A^{\otimes n})=\dsf A-\frac12\,\ell_2(A,A)+\dots\, ,
\end{eqnarray}
then the L$_\infty$ relations with one gauge parameter, ${\cal J}_{n+1}(f, A^{\otimes n})=0$, imply the gauge covariance of such a determined object, 
\begin{eqnarray}\label{FA2}
 \delta_{f} \mathcal{F}= \sum_{n =0}^\infty \, \frac{1}{n!}\,  (-1)^{n(n-1)\over 2}\, \ell_{n+2}(f,\mathcal{F},A,\dots,A)=\ell_2(f,{\cal F})+\ell_3( f,{\cal F},A)+\dots\ .
\end{eqnarray}
To obtain the expression for the corresponding $\ell$-brackets one has to compare these expressions to the corresponding formulas for the Poisson field strength (\ref{pfs}) and the gauge covariance condition. The Poisson field strength (\ref{pfs}) can be expressed as,
\begin{equation}
{\cal F}=\frac12\left(P_{ab}{}^{cd}\left(A\right)\,\partial_c A_d+R_{ab}{}^{cd}\left(A\right)\,\left\{A_c,A_d\right\}\right)\dsf x^a\wedge \dsf x^b\,,\label{4}
\end{equation}
where,
\begin{equation}
P_{ab}{}^{cd}=2\,\gamma^c_lR_{ab}{}^{ld}\,, \qquad\mbox{and}\qquad R_{ab}{}^{cd}=\frac12\left(\rho^c_a\rho^d_b-\rho^c_b\rho^d_a\right)\,.\label{PR}
\end{equation}
The definition (\ref{PR}) and the equation (\ref{second}) imply that the tensors $P_{ab}{}^{cd}$ and  $R_{ab}{}^{cd}$ satisfy the equations \cite{Kup33},
\begin{equation}\label{eqP}
\gamma^k_l\,\partial^l_A P_{ab}{}^{cd}+\Theta^{kl}\,\partial_lP_{ab}{}^{cd}+P_{ab}{}^{cl}\,\partial^d_A\gamma^k_l+P_{ab}{}^{ld}\,\partial_l\Theta^{ck}+2\,R_{ab}{}^{ld}\,\partial_m\gamma^k_l\,\Theta^{mc}=0\,,
\end{equation}
and
\begin{equation}
\gamma^k_l\,\partial^l_A R_{ab}{}^{cd}+\Theta^{kl}\,\partial_lR_{ab}{}^{cd}+R_{ab}{}^{cl}\,\partial^d_A\gamma^k_l+R_{ab}{}^{ld}\,\partial^c_A\gamma^k_l=0\,,\label{eqR}
\end{equation}
correspondingly. We may write,
\begin{eqnarray}\label{coefdef}
&&P_{ab}{}^{cd}(A)=\sum_{n=0}^\infty P_{ab}{}^{cd(n)},\qquad P_{ab}{}^{cd(n)}=P_{ab}{}^{cd|i_1\dots i_n}(x)\,A_{i_1}\dots A_{i_n}\,,\\
&&R_{ab}{}^{cd}(A)=\sum_{n=0}^\infty R_{ab}{}^{cd(n)}\,,\qquad R_{ab}{}^{cd(n)}=R_{ab}{}^{cd|i_1\dots i_n}(x)\,A_{i_1}\dots A_{i_n}\,.\notag
\end{eqnarray}

Comparing (\ref{FA1}) to (\ref{4}) and (\ref{FA2}) to (\ref{gcc}) one writes the following set of non-vanishing brackets,
\begin{eqnarray}\label{An}
\ell_1(A)&=&\dsf A\,,\qquad \ell_2(f, E^A)=-\{f,E^A\}\,,\\
\ell_n\left(A^{\otimes n}\right)&=&\frac{n!}{2}\,(-1)^\frac{n(n-1)}{2}\left(P_{ab}{}^{cd(n-1)}\,\partial_c A_d+R_{ab}{}^{cd(n-2)}\left\{A_c,A_d\right\}\right)\dsf x^a\wedge \dsf x^b\,,\,\,\,n>1\,.\notag
\end{eqnarray}
The off-diagonal elements can be written as,
\begin{eqnarray}\label{Anoff}
\ell_n\left(A^{(1)},\dots,A^{(n)}\right)&=&\frac{(n-1)!}{2}\,(-1)^\frac{n(n-1)}{2}\left[  \sum_{\sigma\in{\rm Sh}_{1,n-1}}\partial_c A^{(\sigma(1))}_d\,P_{ab}{}^{cd|i_2,\dots,i_{n}}A^{(\sigma(2))}_{i_2},\dots,A^{(\sigma(n))}_{i_{n}}\right.\\
&&\left.+\frac{2}{n-1} \sum_{\sigma\in{\rm Sh}_{2,n-2}}\left\{A^{(\sigma(1))}_c,A^{(\sigma(2))}_d\right\}R_{ab}{}^{cd|i_3,\dots,i_{n}}A^{(\sigma(3))}_{i_3},\dots,A^{(\sigma(n))}_{i_{n}}\right]\dsf x^a\wedge \dsf x^b\,.\notag
\end{eqnarray}
Apart from the homotopy relations, ${\cal J}_{n+1}(f, A^{\otimes n})=0$, which will be checked in the Proposition \ref{pr3}, the brackets $\ell_2(f, E^A)=-\{f,E^A\}$ also enters the L$_\infty$ relation with two gauge parameters and one element $E^A\in V_2$, that is, ${\cal J}_{3}(f,g,E^A)=0$. The latter however are trivially satisfied because of the Jacobi identity for the Poisson bracket, just like it happened with, ${\cal J}_{3}(f,g,h)=0$.

One more ingredient of the Poisson gauge theory that we need to express in terms of $\ell$-brackets in the L$_\infty$ formalism is the gauge covariant derivative of matter field (\ref{w6}) which we write as,
\begin{equation}
{\cal D}(\varphi)=\left(\pi_a^i\left(A\right)\partial_i\varphi+\rho_a^i\left(A\right)\left\{A_i,\varphi\right\}\right)\dsf x^a\,,\label{w5}
\end{equation}
where, $\pi_a^i=\rho_a^l\gamma_l^i$, and the coefficient function, 
\be
\rho_{ a }^{ i }(A) =\sum_{n=0}^\infty\rho_{ a }^{ i (n)}\,,\qquad \rho_{ a }^{ i (0)}=\delta_{ a }^{ i }\,,\qquad \rho_{ a }^{ i (n)}=  \rho_{ a }^{ i| j_1\dots j_n }(x) \,A_{j_1}\dots A_{j_n}\,,
\ee 
should satisfy the equation (\ref{second}). In the L$_\infty$ approach one may introduce the covariant derivative as a combination of the brackets involving the matter field and the gauge fields $ \ell_{n+1}(\varphi,A^{\otimes n})$ as,
\begin{eqnarray}\label{Dphi1}
{\cal D}(\varphi)=\sum_{n= 0}   {1\over n!}
      (-1)^{n(n+1)\over 2}\,
\ell_{n+1}(\varphi,A^{\otimes n}) =\dsf \varphi+\ell_2(\varphi,A)+\dots\, .
\end{eqnarray}
Provided that the following brackets vanish, $\ell_{n+3}(f,\varphi, E_A, A^{\otimes n})=0$, which can be assumed here without any contradiction, the homotopy relations, ${\cal J}_{n+2}(f,\varphi,A^{\otimes n})=0$, imply the gauge covariance of the object given by (\ref{Dphi1}), i.e.,
\begin{eqnarray}
\label{varDphi2}
  \delta_{f}  {\cal D}(\varphi) =\sum_{n= 0}^\infty   {1\over n!}
      (-1)^{n(n+1)\over 2}\,
 \ell_{n+2}(f,{\cal D}(\varphi), A^{\otimes n} )=\ell_2(f,{\cal D}(\varphi))- \ell_{3}(f,{\cal D}(\varphi), A)+\dots\,.
\end{eqnarray} 
Comparing (\ref{w5}) to (\ref{Dphi1}) and (\ref{w6}) to (\ref{varDphi2}) we find an additional set of non-vanishing brackets,
\begin{eqnarray}\label{lphiA}
\ell_1(\varphi)&=&\dsf \varphi\,,\qquad \ell_2(f, E^\varphi)=-\{f,E^\varphi\}\,,\\
\ell_{n+1}\left(\varphi,A^{\otimes n}\right)&=&n!\,(-1)^\frac{n(n+1)}{2}\left(\pi_a^{i(n)}\,\partial_i \varphi+\rho_a^{i(n-1)}\,\left\{A_i,\varphi\right\}\right)\dsf x^a\,,\,\,\,n>1\,.\notag
\end{eqnarray}
Just like before, the homotopy relation with two gauge parameters and one element $E^\varphi$ from the subspace $V_2$, ${\cal J}_{3}(f,g,E^\varphi)=0$, is satisfied because of the Jacobi identity for the Poisson bracket. The rest is to prove that the L$_\infty$ relations, ${\cal J}_{n+2}(f, \varphi,A^{\otimes n})=0$, hold true. 

{\proposition\label{pr3} Let the equations (\ref{first}) on the matrix function, $\gamma_k^l(A)=\sum_{n=0}^\infty\gamma_k^{l(n)},$ and (\ref{second}) on, $\rho_a^i(A)=\sum_{n=0}^\infty\rho_a^{i(n)},$ are satisfied. Then there exist a three term L$_\infty$ structure denoted by L$_\infty^{full}$ which induces the Poisson gauge theory \cite{Kup33}. It is concentrated on  $V_{0}\oplus V_{1}\oplus V_{2}$ with, $V_0$, being the space of gauge parameters $f,g\in C^\infty(M)$, the component, $V_1=C^\infty(M)\oplus \Omega^1(M)$, is the space of matter fields $\varphi\in C^\infty(M)$  and gauge fields $A\in\Omega^1(M)$ and $V_2=\Omega^1(M)\oplus\Omega^2(M)$ is the space of the covariant objects $E^\varphi\in\Omega^1(M)$ and $E^A\in\Omega^2(M)$. The diagonal non-vanishing brackets read,
\begin{eqnarray}\label{formulas}
\ell_1(v)&=&\dsf v\,,\qquad\mbox{for}\qquad v=f,A,\varphi\,,\\
\ell_2(f,u)&=&-\{f,u\}\,,\qquad\mbox{for}\qquad u=g,\varphi,E\,,\notag\\
 \ell_2(f,A)&=&\left(\gamma^{k(1)}_a\,\partial_kf+\{A_a,f\}\right)\dsf x^a\,,\notag\\
\ell_{n+1}(f,A^{\otimes n})&=&n! (-1)^{\frac{n\,(n-1)}2}\,\gamma_a^{k(n)}\partial_kf\,\dsf x^a\,,\qquad n\geq2\,.\notag\\
\ell_n\left(A^{\otimes n}\right)&=&\frac{n!}{2}\,(-1)^\frac{n(n-1)}{2}\left(P_{ab}{}^{cd(n-1)}\,\partial_c A_d+R_{ab}{}^{cd(n-2)}\,\left\{A_c,A_d\right\}\right)\dsf x^a\wedge \dsf x^b\,,\,\,\,n>1\,.\notag\\
\ell_{n+1}\left(\varphi,A^{\otimes n}\right)&=&n!\,(-1)^\frac{n(n+1)}{2}\left(\pi_a^{i(n)}\,\partial_i \varphi+\rho_a^{i(n-1)}\,\left\{A_i,\varphi\right\}\right)\dsf x^a\,,\,\,\,n>1\,,\notag
\end{eqnarray}
where
\begin{eqnarray}
R_{ab}{}^{cd}&=&\frac12\left(\rho^c_a\rho^d_b-\rho^c_b\rho^d_a\right)=\sum_{n=0}^\infty R_{ab}{}^{cd(n)}\,,\\
P_{ab}{}^{cd}&=&2\,\gamma^c_lR_{ab}{}^{ld}=\sum_{n=0}^\infty P_{ab}{}^{cd(n)}\,,\qquad\mbox{and}\qquad \pi_a^i=\rho_a^l\gamma_l^i=\sum_{n=0}^\infty \pi_a^{i(n)}\,.\notag
\end{eqnarray}
The graded antisymmetry is imposed by the definition (\ref{eq:gradedantisym}).}

{\it Proof.} To prove it we need to show that all possible homotopy relations involving the above brackets are satisfied. The L$_\infty$ relations with one, two and three entries can be checked explicitly, see \cite{Kup27} and also consideration above in the text concerning, ${\cal J}_3(f,g,u)=0$. Taking into account the consideration of the degree of the brackets the only non-trivial higher relations to be checked are, ${\cal J}_n(f,g,A^{\otimes n-2})=0$,  ${\cal J}_n(f,A^{\otimes n-1})=0$, and ${\cal J}_n(f,\varphi,A^{\otimes n-2})=0$, and were given explicitly by the Proposition \ref{pr2}.

We start with the L$_\infty$ identities ${\cal J}_{n+2}(f,g,A^{\otimes n})=0$\footnote{The original proof can be found in \cite{KS21}, here we provide the proof for completeness}. Let us note that only the bracket $\ell_2(f,A)$ contains the Poisson bracket while $\ell_{m+1}(f,A^{\otimes m})$, with $m>1$ do not. So, it makes sense to write separately all terms in (\ref{Jnfg}) which contain the brackets $\ell_2$. We write separately contributions for $i=2$ and $ i=n+1$.  Taking into account that in the Poisson case $\ell_{n+2}(f,g,A^{\otimes n})=0$ for $n\geq1$, we write the left hand side of ${\cal J}_{n+2}(f,g,A^{\otimes n})=0$ as,
\begin{eqnarray}\label{Jnfg1}
&&(-1)^{n+1}\ell_{n+2}(\ell_1(f),g,A^{\otimes n})+(-1)^{n+1}\ell_{n+2}(f,\ell_1(g),A^{\otimes n})+\ell_{n+1}\left(\ell_i(f,g),A^{\otimes n}\right)\\
&&-n\,\ell_{n+1}\left(\ell_2(f,A),g,A^{\otimes n-1}\right)+n\,\ell_{n+1}\left(\ell_2(g,A),f,A^{\otimes n-1}\right)+  \notag\\
&& -\ell_{2}\left(\ell_{n+1}(f,A^{\otimes n}),g\right)+\ell_{2}\left(\ell_{n+1}(g,A^{\otimes n}),f\right) \notag\\
&&+\sum_{i=3}^{n}(-1)^{i(n-i)+i-1}\mbox{$\binom{n}{i-1}$}\left[\ell_{n-i+3}\left(\ell_i(f,A^{\otimes i-1}),g,A^{\otimes n-i+1}\right)-\ell_{n-i+3}\left(\ell_i(g,A^{\otimes i-1}),f,A^{\otimes n-i+1}\right)\right]\,. \notag
\end{eqnarray}
We substitute (\ref{h6}) in (\ref{Jnfg1}) and after careful simplification one may see that all terms containing the Poisson brackets like $\{f,A_m\}$ or $\{\partial_m f,g\}$ will cancel each other. The rest of the terms can be reorganized as (in particular we set $i=m+1$ in the sum),
\begin{eqnarray}\label{h9}
{\cal J}_{n+2}(f,g,A^{\otimes n})&=&n!\,(-1)^{\frac{n(n-1)}{2}}\left[(n+1)\left(\gamma_i^{k|li_1\dots i_{n}}-\gamma_i^{l|ki_1\dots i_{n}}\right)+\gamma_j^{l|i_1\dots i_{n}}\,\gamma_i^{k|j} -
\gamma_j^{k|i_1\dots i_{n}}\,\gamma_i^{l|j}\right.+\\
&& -\Theta^{kj}\,\partial_j\gamma_i^{l|i_1\dots i_{n}} + \Theta^{lj}\,\partial_j\gamma_i^{k|i_1\dots i_{n}} -
\gamma_i^{j|i_1\dots i_{n}}\,\partial_j\Theta^{lk}+\notag\\
&&\sum_{m=1}^{n-1}\left.(n-m+1)\left(\gamma_j^{l|i_1\dots i_{m}}\,\gamma_i^{k|ji_{m+1}\dots i_{n}} -
\gamma_j^{k|i_1\dots i_{m}}\,\gamma_i^{l|ji_{m+1}\dots i_{n}} \right)\right]\partial_kf\,\partial_lg\,A_{i_1}\dots A_{i_{n}}\,. \notag
\end{eqnarray}
On the other hand, substituting the decomposition (\ref{h3}) in the equation (\ref{first}) one obtains,
\begin{eqnarray}\label{h5}
&&\gamma^{k|l}_i-\gamma^{l|k}_i-\partial_i\Theta^{lk} +\\
&&\left[2\left(\gamma_i^{k|li_1} -\gamma_i^{l|ki_1}\right)+ \gamma_{ j }^{ l| i_1 }\gamma_{ i }^{ k|j }-\gamma_{ j }^{ k| i_1 }\gamma_{ i }^{ l|j }-\Theta^{kj}\,\partial_j\gamma_i^{l|i_1} + \Theta^{lj}\,\partial_j\gamma_i^{k|i_1} -
\gamma_i^{j|i_1}\,\partial_j\Theta^{lk}  \right]A_{i_1}+\notag\\
&&\sum_{n=2}^\infty\left[(n+1)\left(\gamma_i^{k|li_1\dots i_{n}}-\gamma_i^{l|ki_1\dots i_{n}}\right)+\gamma_j^{l|i_1\dots i_{n}}\,\gamma_i^{k|j} -
\gamma_j^{k|i_1\dots i_{n}}\,\gamma_i^{l|j}\right.+\notag\\
&& -\Theta^{kj}\,\partial_j\gamma_i^{l|i_1\dots i_{n}} + \Theta^{lj}\,\partial_j\gamma_i^{k|i_1\dots i_{n}} -
\gamma_i^{j|i_1\dots i_{n}}\,\partial_j\Theta^{lk}+\notag\\
&&\sum_{m=1}^{n-1}\left.(n-m+1)\left(\gamma_j^{l|i_1\dots i_{m}}\,\gamma_i^{k|ji_{m+1}\dots i_{n}} -
\gamma_j^{k|i_1\dots i_{m}}\,\gamma_i^{l|ji_{m+1}\dots i_{n}} \right)\right]A_{i_1}\dots A_{i_{n}}=0\,. \notag
\end{eqnarray}
The above relation should hold for any gauge field $A$ so we get separately,
\begin{eqnarray*}\label{h5_0}
&&\gamma^{k|l}_i-\gamma^{l|k}_i-\partial_i\Theta^{lk} =0\,,\\
&&\left[2\left(\gamma_i^{k|li_1} -\gamma_i^{l|ki_1}\right)+ \gamma_{ j }^{ l| i_1 }\gamma_{ i }^{ k|j }-\gamma_{ j }^{ k| i_1 }\gamma_{ i }^{ l|j }-\Theta^{kj}\,\partial_j\gamma_i^{l|i_1} + \Theta^{lj}\,\partial_j\gamma_i^{k|i_1} -
\gamma_i^{j|i_1}\,\partial_j\Theta^{lk}  \right]A_{i_1}=0\,,\label{h5_1}\\
&&\sum_{n=2}^\infty\left[(n+1)\left(\gamma_i^{k|li_1\dots i_{n}}-\gamma_i^{l|ki_1\dots i_{n}}\right)+\gamma_j^{l|i_1\dots i_{n}}\,\gamma_i^{k|j} -
\gamma_j^{k|i_1\dots i_{n}}\,\gamma_i^{l|j}\right.+\label{h5_n}\\
&& -\Theta^{kj}\,\partial_j\gamma_i^{l|i_1\dots i_{n}} + \Theta^{lj}\,\partial_j\gamma_i^{k|i_1\dots i_{n}} -
\gamma_i^{j|i_1\dots i_{n}}\,\partial_j\Theta^{lk}+\notag\\
&&\sum_{m=1}^{n-1}\left.(n-m+1)\left(\gamma_j^{l|i_1\dots i_{m}}\,\gamma_i^{k|ji_{m+1}\dots i_{n}} -
\gamma_j^{k|i_1\dots i_{m}}\,\gamma_i^{l|ji_{m+1}\dots i_{n}} \right)\right]A_{i_1}\dots A_{i_{n}}=0\,. \notag
\end{eqnarray*}
Which implies that the right hand side of (\ref{h9}) vanishes. Thus, the fact that the matrix function $\gamma_k^l(A)$ satisfies the equation (\ref{first}) implies that the brackets (\ref{h6}) indeed satisfy the L$_\infty$ relations (\ref{Jnfg}).

We follow the same logic to prove that the $\ell$-brackets defined in (\ref{An}) satisfy the L$_\infty$ relations, ${\cal J}_{n+1}(f, A^{\otimes n})=0$, given by (\ref{JnfA}). Since the only non-vanishing bracket containing an element $E_A\in V_2$ is $\ell_2(f,E_A)=\{f,E_A\}$, the only contribution from the fourth line of (\ref{JnfA}) appears when $i=n-1$. So for $n>2$ we have,
\begin{eqnarray}\label{jna}
{\cal J}_{n+1}(f, A^{\otimes n})&=&(-1)^n\ell_{n+1}\left(\ell_1(f),A^{\otimes (n)}\right)+\ell_2\left(\ell_n\left(A^{\otimes(n)}\right),f\right)+n\,\ell_n\left(\ell_2(f,A),A^{(n-1)}\right)+\\
&&\sum_{i=3}^n (-1)^{i(n-i+1)}\mbox{$\binom{n}{i-1}$}\ell_{n-i+2}\left(\ell_i\left(f,A^{\otimes(i-1)}\right),A^{\otimes(n-i+1)}\right)+\ell_1\left(\ell_{n+1}\left(f,A^{\otimes(n)}\right)\right)\,.\notag
\end{eqnarray}
To check (\ref{JnfA}) we will need the expression for off-diagonal elements of the bracket $\ell_n(A^{\otimes n})$ given by (\ref{Anoff}). In particular,
\begin{eqnarray}
\ell_n\left(B,A^{\otimes(n-1)}\right)&=&(n-1)!\,(-1)^\frac{n(n-1)}{2}\left((n-1)\,P_{ab}{}^{cd(n-1)}(B,A)\,\partial_c A_d+P_{ab}{}^{cd(n-1)}(A)\,\partial_c B_d\right.\notag\\
&&\left.+(n-2)R_{ab}{}^{cd(n-2)}\left(B,A\right)\,\left\{A_c,A_d\right\}+2\,R_{ab}{}^{cd(n-2)}\left(A\right)\,\left\{B_c,A_d\right\}\right)\,,\label{lBA}
\end{eqnarray}
where we use the notation,
\begin{equation}
P_{ab}{}^{cd(k)}(B,A)=P_{ab}{}^{cd|i_1\dots i_k}B_{i_1}A_{i_2}\dots A_{i_k}\,.
\end{equation}
We use (\ref{lBA}) in (\ref{jna}) with, $B=\ell_{k+1}(f,A^{\otimes k})$, given by (\ref{An}). The resulting expression is huge, but different simplifications can be made. In particular, all terms containing double Poisson brackets, like $R_{ab}{}^{cd(n-2)}\{f,\{A_c,A_d\}\}$, vanish because of the Jacobi identity for the Poisson bracket. Some other terms cancel each other. Such that the resulting expression after the careful simplification becomes,
\begin{eqnarray}\label{JnA1}
{\cal J}_{n+1}(f, A^{\otimes n})&=&n!\,(-1)^{\frac{n(n-1)}{2}}\left\{{\cal U}_{ab}{}^{cdk(n-1)}\partial_cA_d\,\partial_kf+\sum_{m=0}^nP_{ab}{}^{cd(m)}\,\gamma_d^{k(n-m)}\partial_c\partial_kf+\right.\\
&&\left.\left[P_{ab}{}^{cd(n-1)}-2\sum_{m=0}^{n-1}\gamma_l^{c(m)}\,R_{ab}{}^{ld(n-m-1)} \right]\{A_d,\partial_cf\}+{\cal V}_{ab}{}^{cdk(n-2)}\{A_c,A_d\}\,\partial_kf\right\}\,.\notag
\end{eqnarray}
where the coefficients ${\cal U}_{ab}{}^{cdk(n)}$ and ${\cal V}_{ab}{}^{cdk(n)}$ are given by,
\begin{eqnarray}
{\cal U}_{ab}{}^{cdk(0)}&=&P_{ab}{}^{cd|k}+\delta_a^c\,\gamma_b^{k|d}-\delta_b^c\,\gamma_a^{k|d}+\delta_b^d\,\partial_a\Theta^{ck}-\delta_a^d\,\partial_b\Theta^{ck}\,,\\
{\cal U}_{ab}{}^{cdk(1)}&=&\left[2\,P_{ab}{}^{cd|ki_1}+\gamma_l^{k|i_1}P_{ab}{}^{cd|l}+\Theta^{kl}\partial_lP_{ab}{}^{cd|i_1}+2\,\delta_a^c\,\gamma_b^{k|di_1}-2\,\delta_b^c\,\gamma_a^{k|di_1}+\right.\notag\\
&&\left.P_{ab}{}^{cl|i_1}\gamma_l^{k|d}+P_{ab}{}^{ld|i_1}\partial_l\Theta^{ck}+\delta_b^d\partial_q\gamma_a^{k|i_1}\Theta^{qc}-\delta_a^d\partial_q\gamma_b^{k|i_1}\Theta^{qc}\right]A_{i_1}\,,\notag\\
{\cal U}_{ab}{}^{cdk(n)}&=&\left[(n+1)\,P_{ab}{}^{cd|ki_1\dots i_n}+\gamma_l^{k|i_1\dots i_n}P_{ab}{}^{cd|l}+\Theta^{kl}\partial_lP_{ab}{}^{cd|i_1\dots i_n}+\right.\notag\\
&&(n+1)\,\delta_a^c\,\gamma_b^{k|di_1\dots i_n}-(n+1)\,\delta_b^c\,\gamma_a^{k|di_1\dots i_n}+P_{ab}{}^{cl|i_1\dots i_n}\gamma_l^{k|d}+\notag\\
&&P_{ab}{}^{ld|i_1\dots i_n}\partial_l\Theta^{ck}+\delta_b^d\partial_q\gamma_a^{k|i_1\dots i_n}\Theta^{qc}-\delta_a^d\partial_q\gamma_b^{k|i_1\dots i_n}\Theta^{qc}+\notag\\
&&\sum_{m=1}^{n-1}\left((n-m+1)\,\gamma_l^{k|i_1\dots i_m}R_{ab}{}^{cd|li_{m+1}\dots i_n}+(n-m+1)\,P_{ab}{}^{cl|i_1\dots i_m}\gamma_l^{k|di_{m+1}\dots i_n}\right.\notag\\
&&\left.\left.+2\,R_{ab}{}^{ld|i_1\dots i_m}\partial_q\gamma_l^{k|i_{m+1}\dots i_n}\Theta^{qc}\right)\right] A_{i_1}\dots A_{i_n}\,.\notag
\end{eqnarray}
and
\begin{eqnarray}
{\cal V}_{ab}{}^{cdk(0)}&=&R_{ab}{}^{cd|k}+\frac12\left(\delta_a^c\,\gamma_b^{k|d}-\delta_b^c\,\gamma_a^{k|d}+\delta_b^d\,\gamma_a^{k|c}-\delta_a^d\,\gamma_b^{k|c}\right)\,,\\
{\cal V}_{ab}{}^{cdk(1)}&=&\left[2\,R_{ab}{}^{cd|ki_1}+\gamma_l^{k|i_1}\,R_{ab}{}^{cd|l}+\Theta^{kl}\,\partial_lR_{ab}{}^{cd|i_1}+\right.\notag\\
&&\left.\delta_a^c\,\gamma_b^{k|di_1}-\delta_b^c\,\gamma_a^{k|di_1}+\delta_b^d\,\gamma_a^{k|ci_1}-\delta_a^d\,\gamma_b^{k|ci_1}\right]A_{i_1}\,,\notag\\
{\cal V}_{ab}{}^{cdk(n)}&=&\left[(n+1)\,R_{ab}{}^{cd|ki_1\dots i_n}+\gamma_l^{k|i_1\dots i_n}\,R_{ab}{}^{cd|l}+\Theta^{kl}\,\partial_lR_{ab}{}^{cd|i_1\dots i_n}+\right.\notag\\
&&\frac{n+1}{2}\left(\delta_a^c\,\gamma_b^{k|di_1\dots i_n}-\delta_b^c\,\gamma_a^{k|di_1\dots i_n}+\delta_b^d\,\gamma_a^{k|ci_1\dots i_n}-\delta_a^d\,\gamma_b^{k|ci_1\dots i_n}\right)+\notag\\
&&R_{ab}{}^{cl|i_1\dots i_n}\,\gamma^{k|d}_l+R_{ab}{}^{ld|i_1\dots i_n}\,\gamma^{k|c}_l+\notag\\
&&\sum_{m=1}^{n-1}(n-m+1)\left(\gamma_l^{k|i_1\dots i_m}R_{ab}{}^{cd|li_{m+1}\dots i_n}+R_{ab}{}^{cl|i_1\dots i_m}\,\gamma^{k|di_{m+1}\dots i_n}_l\right.\notag\\
&&\left.\left.+R_{ab}{}^{ld|i_1\dots i_m}\,\gamma^{k|ci_{m+1}\dots i_n}_l\right)\right] A_{i_1}\dots A_{i_n}\,.\notag
\end{eqnarray}
Now we observe that substituting the decompositions for $R_{ab}{}^{cd}(A)$ and $\gamma_k^l(A)$ in the equation (\ref{eqR}) one may represent it as,
\begin{equation}\label{eqRdec}
\sum_{n=0}^\infty {\cal V}_{ab}{}^{cdk(n)}=0\,.
\end{equation}
Just like it happened with the equation (\ref{first}) in (\ref{h5}).
Each term ${\cal V}_{ab}{}^{cdk(n)}$ is of the order $n$ in the gauge fields $A_a$. Since (\ref{eqRdec}) should hold for any gauge field $A_a$, it implies that,
\begin{equation}
{\cal V}_{ab}{}^{cdk(n)}=0\,,\qquad n=0,1,\dots
\end{equation}
The term in the square brackets in the second line of (\ref{JnA1}) is nothing but the first of the eqs. (\ref{PR}) written in components, therefore it also vanishes. The second term in the first line of (\ref{JnA1}) vanishes as a consequence of (\ref{PR}) and the antisymmetry of the coefficient function $R_{ab}{}^{cd}(A)$ in the upper two indices. Finally, it was shown in \cite{Kupriyanov:2020sgx} that the equations (\ref{eqR}) and (\ref{first}) imply that the coefficient function $P_{ab}{}^{cd}(A)$ defined by (\ref{PR}) satisfies the equation (\ref{eqP}).
The left hand side of (\ref{eqP}) can be written as $\sum_{n=0}^\infty {\cal U}_{ab}{}^{cdk(n)}$, from where one finds,
\begin{equation}
{\cal U}_{ab}{}^{cdk(n)}=0\,,\qquad n=0,1,\dots\,.
\end{equation}
Thus we conclude that right hand side of (\ref{JnA1}) vanishes and consequently the L$_\infty$ relations, ${\cal J}_{n+1}(f, A^{\otimes n})=0$ hold true. 

The fact that the $\ell$-brackets defined in (\ref{lphiA}) satisfy the relations, ${\cal J}_{n+1}(f, A^{\otimes n})=0$, given explicitly in (\ref{Jnfphi}), can be proved in the absolutely the same way, using the decompositions of the equation (\ref{second}) and the equation on the matrix function, $\pi_a^i=\rho_a^l\gamma_l^i$. $\Box$

 Now let us turn to the Maxwell-Poisson equations of motion. Aiming to arrive at the cyclic L$_\infty$ algebras we work with the equations (\ref{EL1}) obtained from the action (\ref{ag}). 
Following \cite{Kup33} one may check that,
\begin{eqnarray}
\delta S_g=\int dx^n \,\delta A_a\cdot{\cal E}^a_{EL}(A)\,,
\end{eqnarray}
where, ${\cal E}^a_{EL}(A)$ is given by (\ref{EL}). Comparing this formula with the corresponding expression in the L$_\infty$ formalism \cite{Hohm:2017pnh}, 
\begin{equation}
\delta S=\langle\delta A,E_A\rangle\,,
\end{equation}
  we identify the inner product on the L$_\infty$ algebra as,
 \begin{equation}
\langle A,E\rangle=\int dx^n \,A_a\,E^a\,,
\end{equation}
and the corresponding field equations become,
  \begin{eqnarray}\label{FEL}
{\cal E}_{EL}(A)=\sum^\infty_{n=1} \;{1\over n!}
(-1)^\frac{n(n-1)}{2}\, \ell^{\cal E}_n(A^{\otimes n})=\ell_1^{\cal E} (A)-\frac12\,\ell^{\cal E}_2(A,A)+\dots\, .
\end{eqnarray}
In order to obtain explicit expressions for the brackets $ \ell^{\cal E}_n(A^{\otimes n})$ we rewrite the Maxwell-Poisson equations as follows:
 \be
{\cal E}^{q}_{EL}(A) =  -\frac{\mu}{2}\, \frac{\partial \mathcal{F}_{ab}}{\partial A_q}\,\mathcal{F}^{ab} 
+\frac{1}{2}\,\partial_r\left(\mu\, \frac{\partial \mathcal{F}_{ab}}{\partial(\partial_r A_q)}\,\mathcal{F}^{ab}\right). \la{ourPsi}
\ee
Using the expression~\eqref{4} for the field strength,
one obtains the expansions:
\be
\mathcal{F}_{ab} = \sum_{n=1}^{\infty}\mathcal{F}_{ab}^{(n)},\quad \quad \frac{\partial \mathcal{F}_{ab}}{\partial A_q} = \sum_{n=0}^{\infty} \mathcal{S}_{ab}^{q\, (n)},\quad\quad 
\frac{\partial \mathcal{F}_{ab}}{\partial(\partial_r A_q)} = \sum_{n=0}^{\infty} \mathcal{Q}_{ab}^{rq\, (n)}, \la{Fexpe}
\ee
where the quantities 
\bea
\mathcal{F}^{(1)}_{ab} &:=&  
 \partial_a A_b - \partial_b A_a,
\nonumber\\
\mathcal{F}^{(n)}_{ab} &:=&  
 P_{ab~~}^{~~cd| i_1 i_2 i_3... i_{n-1}} A_{i_1}...A_{i_{n-1}} \,\partial_c A_d 
 +R_{ab~~}^{~~cd|i_1 i_2 i_3... i_{n-2}} A_{i_1}...A_{i_{n-2}} \{A_c,A_d\} ,\quad n\geq 2,
 \nonumber\\
  \mathcal{S}_{ab}^{q\, (0)} &:=& 0,\nonumber\\
 \mathcal{S}_{ab}^{q\, (1)} &:=& P_{ab~~}^{~~cd|q} \,\partial_c A_d,\nonumber\\
\mathcal{S}_{ab}^{q\, (n)} &:=& n\,P_{ab~~}^{~~cd|q\, i_2 i_3... i_{n}} A_{i_2}...A_{i_{n}} \,\partial_c A_d
+ (n-1)\, R_{ab~~}^{~~cd|q\, i_2 i_3... i_{n-1}} A_{i_2}...A_{i_{n-1}} \{A_c,A_d\},\quad n\geq2 \nonumber\\
\mathcal{Q}_{ab}^{rq\, (0)} &:=& \delta_a^r\delta_b^q - \delta_b^r\delta_a^q, \nonumber\\
\mathcal{Q}_{ab}^{rq\, (n)} &:=&  P_{ab~~}^{~~rq| i_1 i_2 i_3... i_{n}} A_{i_1}...A_{i_{n}}   + 2 R_{ab~~}^{~~cd|i_1 i_2 i_3... i_{n-1}} A_{i_1}...A_{i_{n-1}}   \,\Theta^{pr} \,\partial_{p}A_c, \quad n\geq 1 \la{interm1}
\eea
are $n$-linear in $A$. We remind that for any $n\in \mathbb{N}$ the coefficient functions $P_{ab~~}^{~~cd| i_1 i_2 i_3... i_{n}}$ and $R_{ab~~}^{~~cd|i_1 i_2 i_3... i_{n}} $ are defined in Eq.~\eqref{coefdef}.

Substituting the expansions~\eqref{interm1} in the equations of motion~\eqref{ourPsi}, we obtain
\be
{\cal E}^{q}_{EL}(A) = \sum_{n=0}^{\infty} {\cal E}^{q\,(n)}_{EL}, \label{FE2}
\ee
where the combinations
\be
{\cal E}^{q\,(n)}_{EL} = \sum_{k=0}^{n-1} \bigg[ -\frac{\mu}{2}\, \mathcal{S}_{ab}^{q\, (k)}  \,\mathcal{F}^{ab (n-k)} 
+\frac{1}{2}\,\partial_r\left(\mu\, \mathcal{Q}_{ab}^{rq\, (k)}\,\mathcal{F}^{ab\,(n-k)}\right)  \bigg] 
\ee
are $n$-linear in $A$.

Confronting the decompositions~\eqref{FEL} and~\eqref{FE2} we obtain the required brackets:
\be\label{lEn}
\ell^{\cal E}_n(A^{\otimes n}) = n! \,(-1)^\frac{n(n-1)}{2} \,\sum_{k=0}^{n-1} \bigg[ -\frac{\mu}{2}\, \mathcal{S}_{ab}^{q\, (k)}  \,\mathcal{F}^{ab (n-k)} 
+\frac{1}{2}\,\partial_r\left(\mu\, \mathcal{Q}_{ab}^{rq\, (k)}\,\mathcal{F}^{ab\,(n-k)}\right)  \bigg]\dd x^q, 
\ee
where the quantities $\mathcal{F}^{(n)}_{ab}$, $\mathcal{S}_{ab}^{q\, (n)}$, and $\mathcal{S}_{ab}^{q\, (n)}$ are defined in Eq.~\eqref{interm1} in terms of the coefficient functions 
$P_{ab~~}^{~~cd| i_1 i_2 i_3... i_{n}}$ and $R_{ab~~}^{~~cd|i_1 i_2 i_3... i_{n}} $, associated with the given Poisson gauge theory.
The fact that the action is gauge invariant, $\delta_fS_g\equiv0$, implies that the solutions of the field equations (\ref{FEL}) are mapped onto the solutions under the gauge transformations. In this sense the equations (\ref{FEL}) are gauge covariant.  The latter can be translated to the fact that the brackets (\ref{lEn}) satisfy the corresponding homotopy relations just like it was shown explicitly in the Proposition \ref{pr3} for the brackets (\ref{An}). 

As a final remark we mention here that the important identities involving the covariant objects from the space $V_2$, like the Bianchi identity for the field strength, the commutator relation for the covariant derivatives or the Noether identities for the field equations, also can be obtained in the L$_\infty$ formalism. However to this end we will need an additional non-empty subspace of degree three - the space of identities. Then, from the non-vanishing brackets of the form, $ \ell_{n+1}({\cal F},A^{\otimes n})\in V_3$, one may construct the object, 
\begin{align}\label{Bianchi1}
\dsf_A{\cal F} = \sum_{n= 0}^\infty   {1\over n!}
      (-1)^{n(n-1)\over 2}\,
 \ell_{n+1}({\cal F},A^{\otimes n})\ .
\end{align}
The L$_\infty$ relations, ${\cal J}_n(A^{\otimes n})=0$, imply that, $\dsf_A{\cal F} \equiv0$, that is we have the Bianchi identity for the field strength ${\cal F}$, see \cite{Szabo1,Jurco:2018sby} for details. We leave the detailed analysis of these identities and the corresponding L$_\infty$ structures for the future research.

\section{Examples.}
In this section we illustrate our results, presenting explicit L$_{\infty}$ structures for the particular choices of non-commutativity parameter, namely the $\kappa$-Minkowski and the $\mathfrak{su}(2)$ structures. For the sake of simplicity we abandon the matter fields, considering the gauge fields only.  The brackets of the structure $\ell_n\left(f,A^{\otimes n-1}\right)$, $f\in V_0$, $A\in V_1$ have already been studied in~\cite{KS21}, therefore  all the remaining nontrivial (i.e. noncommutativity-dependent)  brackets have the structure
$\ell_n\left(A^{\otimes n}\right)$, $n>1$.

Our technical strategy is as follows. Starting from the known solutions for $\gamma$ and $\rho$, we expand the quantity $\mathcal{F}_{ab}$ defined by (\ref{pfs})
in powers  of $A$,
\be
\mathcal{F}_{ab} = \sum_{n=0}^{\infty} \mathcal{F}_{ab}^{(n)}, \la{Fexp}
\ee
where, by definition, $\mathcal{F}_{ab}^{(n)}$ is $n$-linear\footnote{i.e. upon the rescaling $A \to \Omega A$, $\Omega\in\mathbb{R}$, the quantity $\mathcal{F}_{ab}^{(n)}$ rescales as 
$\mathcal{F}_{ab}^{(n)} \to \Omega^n \mathcal{F}_{ab}^{(n)}$} in $A$.  It is worth noticing that the combination $$ P_{ab}{}^{cd(n-1)}\,\partial_c A_d+R_{ab}{}^{cd(n-2)}\,\left\{A_c,A_d\right\}, $$ which enters in the Proposition~\ref{pr3} is nothing but $\mathcal{F}_{ab}^{(n)}$.
Thus we obtain explicit expressions for the $\ell_n\left(A^{\otimes n}\right)$-brackets in the following form
\be
\ell_n\left(A^{\otimes n}\right)= \frac{n!}{2}\,(-1)^\frac{n(n-1)}{2}\,\mathcal{F}_{ab}^{(n)}\,\dsf x^a\wedge \dsf x^b . \la{linfF}
\ee

\subsection*{a. $\kappa$-Minkowski.}
In the $N$-dimensional $\kappa$-Minkowski case the Poisson bi-vector is given by 
\be
\Theta^{ij} = 2(\omega^ix^j - \omega^j x^i),
\ee
where $\omega^i$, $i=1,...,N$ are deformation parameters\footnote{In the Euclidean case $\omega^i$ can be transformed by a linear transformation of the coordinates (rotation) to the form:
$\omega^i = \delta^{iN}\kappa^{-2}$, see \cite{Wess}.}.
The corresponding structure constants read:
\be
f^{aj}_k  = 2(\omega^a\delta^j_k - \omega^j\delta^a_k).
\ee
For the forthcoming discussion it is convenient to introduce the quantity
\be
z := \omega^i A_i.
\ee 
The matrices $\gamma$ and $\rho$ are given by~\cite{Kup33,KKV}:
\be
\gamma^i_a = \delta^i_a\,\xi(z) - \omega^i A_a,
\ee
and
\be
\rho^i_a = \delta^i_a \,\xi(z) - \omega^iA_a \,\xi'(z),
\ee
respectively, with $\xi'(z)=d\xi/dz$. In these formulae the form factor $\xi$ is defined as follows:
\be
\xi(z) := \sqrt{1+z^2} + z.
\ee
Substituting this data in our general formula~\eqref{pfs},
we arrive at\footnote{This formula is not new, it has already appeared in~\cite{KKV}. However, the present parametrisation is the most convenient for the forthcoming expansion in powers of $A$.}
\bea
\mathcal{F}_{ab} &=& \big(\delta^j_b \,\xi(z) - \omega^jA_b \,\xi'(z)\big)\, \partial_a A_j \nonumber\\
 &+& \left( A_q \partial_l A_j\, f_i^{ql} + \frac{1}{2}\{A_i,A_j\}\right)\Big( \delta_a^i\delta_b^j \cdot\big(2 \,z\, \xi(z) + 1\big)
 - \big(\delta_a^i\omega^j A_b + \delta_b^j\omega^iA_a\big)\cdot \big(z\,\xi(z)\big)' \Big)  \nonumber\\
 &-& \Big(a\leftrightarrow b\Big) . \la{Fkappa}
\eea
The form factor $\xi$ exhibits the following Taylor expansion in powers of $z$:
\be
\xi(z) = \sum_{n=0}^{\infty} \xi_n\,z^n,
\ee
where
\be
\xi_n = \left\{
\begin{array}{l}
\frac{(-1)^{m-1} (2m)!}{4^m (m!)^2 (2m-1)}, \quad n = 2m, \quad m\in\mathbb{Z}_+,\\
1, \quad n=1,\\
0, \quad n = 2m+1,\quad m\in\mathbb{N}.
\end{array}
\right. \la{xindef}
\ee
Substituting this expansion in Eq.~\eqref{Fkappa}, and taking into account its obvious consequences,
\bea
2 \,z\, \xi(z) + 1 &=& 1 + \sum_{n=1}^{\infty} (2\,\xi_{n-1})\, z^n,\nonumber\\
\big(z\,\xi(z)\big)' &=& \sum_{n=0}^{\infty} \big((n+1)\,\xi_n\big)\, z^n,
\eea
we obtain the expansion~\eqref{Fexp} with
\bea
\mathcal{F}_{ab}^{(1)} &=& \partial_a A_b  - \partial_b A_a, \nonumber\\
\mathcal{F}_{ab}^{(2)} &=& \big(z\,\delta_b^j - \omega^j A_b\big)\, \partial_a A_j 
+  A_q\partial_l A_b\,f_a^{ql}  + \frac{1}{2}\{A_a,A_b\} - (a\leftrightarrow b), \nonumber\\
\mathcal{F}_{ab}^{(n)} &=& \big(z\,\delta_b^j - (n-1)\, \omega^j A_b\big)\, \partial_a A_j \,z^{n-2}\,\xi_{n-1} +
\left( A_q\partial_l A_j\,f_{i}^{ql}  + \frac{1}{2}\{A_i,A_j\} \right)\times \nonumber\\
&\times&  \big[ 2\,\delta_a^i\delta_b^j\, z - (n-2)(\delta_a^i \omega^j A_b + \delta_b^j\omega^i A_a) \big]\, z^{n-3} \xi_{n-3} 
- (a\leftrightarrow b), \quad n\geq3.
\eea
Therefore, according to Eq.~\eqref{linfF}, we arrive at the required L$_{\infty}$ brackets:
\bea
\ell_2\left(A^{\otimes 2}\right)&=& -2 \Big[  \big(z\,\delta_b^j - \omega^j A_b\big)\, \partial_a A_j 
+  A_q\partial_l A_b\,f_a^{ql}  + \frac{1}{2}\{A_a,A_b\} \Big]\,\dsf x^a\wedge \dsf x^b, \nonumber\\
\ell_n\left(A^{\otimes n}\right)&=& {n!}\,(-1)^\frac{n(n-1)}{2}\, \Big[ \big(z\,\delta_b^j - (n-1)\, \omega^j A_b\big)\, \partial_a A_j \,z^{n-2}\,\xi_{n-1}+\left( A_q\partial_l A_j\,f_{i}^{ql}  + \frac{1}{2}\{A_i,A_j\} \right)\times \nonumber\\
&\times&  \big[ 2\,\delta_a^i\delta_b^j\, z - (n-2)(\delta_a^i \omega^j A_b + \delta_b^j\omega^i A_a) \big]\, z^{n-3} \xi_{n-3} 
\Big]\,\dsf x^a\wedge \dsf x^b, \quad n\geq 3,
\eea
where, we remind, the coefficients $\xi_n$ are defined by Eq.~\eqref{xindef}.

\subsection*{b. $\mathfrak{su}(2)$-structure.}
For the $\mathfrak{su}(2)$-case,
\be
\Theta^{jk} = f^{jk}_{l}\,x^l,\quad \quad \varepsilon^{j k}_{~~\,l}:= \varepsilon^{j k s}\,\delta_{s l}, 
\ee
and hence,
\be
f^{jk}_{l}  =  2\lambda\,\varepsilon^{j k}_{~~\,l},  \la{su2}
\ee
where $\lambda$ is a deformation parameter.
Solutions of the master equations can be chosen as follows~\cite{Kupriyanov:2020sgx,Kup33}:
 \be
 \gamma^k_a = \big[1 + {Z\,} \chi(Z)\big]\delta_a^k - \lambda^2 \chi(Z) \,A_a A^k - \lambda\, \varepsilon^{kl}_{~~\,a} A_l , \quad\quad Z :=  \lambda^2\,A_{ i } A^{ i },\quad\quad A^{ j} := \delta^{ j\xi} A_{\xi}
 \ee
and
\be
\rho^i_a = {\phi}(Z)\,\delta_a^i + \lambda^2\tau(Z)A^i A_a {-} \lambda\,{\beta}(Z) \varepsilon^{ik}_{~~\,a} A_k, \la{su2soluRho}
\ee
 where, by definition,
 \bea
 \chi(t) &=& \frac{1}{t}\cdot \big(\sqrt{t}\cot{\sqrt{t}} - 1\big) \nonumber\\
 {\phi}(t) &=& \frac{\sin{2\sqrt{t}}}{2\sqrt{t}} \nonumber \\
 \tau(t) &=& -\frac{1}{t}\cdot\bigg( \frac{\sin{2\sqrt{t}}}{2\sqrt{t}} - 1 \bigg) \nonumber\\
{\beta(t) }&{=}&{ \frac{ \big(\sin{\sqrt{t}}\big)^2  }{t}}.
 \eea
 Substituting these expressions in Eq.~\eqref{4}, and performing a few algebraic simplifications, we arrive at the know expression for the deformed field strength~\cite{Kupriyanov:2020sgx}:
\be
\mathcal{F}_{ab} = \partial_c\pi_{ab}^{~~\,c} + 2\,\big\{\rho_{ab}^{~~\,c} ,A_c\big\}, \la{Fsu2}
\ee 
with 
\bea
\pi_{ab}^{~~\,c} &:=& \delta_{a}^c\, A_b \, \phi(Z) - \frac{\lambda}{2}\,\varepsilon_{abm} \,A^m\, A^c \,\beta(Z) + \frac{1}{2\lambda}\, \varepsilon_{ab}^{~~\,c}\,\Phi(Z) 
-(a\leftrightarrow b),\nonumber\\
\rho_{ab}^{~~\,c} &:=& \frac{1}{4}\,\delta_b^c \,A_a\,\beta(Z) - \frac{1}{8\lambda}\,\varepsilon_{ab}^{~~\, c}\,\Lambda(Z)  - (a\leftrightarrow b). \la{phirhodef}
\eea
In this formulae the form factors {$\Lambda$ and $\Phi$} are defined in the following way:
\bea
\Lambda(t) &=& \int_0^{t} \dd u\, \beta(u), \nonumber\\
\Phi(t) &=&   \int_0^{t} \dd u\, \phi(u) = t\cdot\beta(t).
\eea
Substituting the Taylor series 
\bea
\beta(t)    &=& \sum_{m=0}^{\infty} \frac{(-1)^m\, 4^{m+\frac{1}{2}}}{(2\,m+2)!}\,t^m, \quad\quad 
\Lambda(t) = \sum_{m=1}^{\infty} \frac{(-1)^{m+1} \,4^{m-\frac{1}{2}}}{(2\,m)!\,m}\,t^m,\nonumber\\
\phi(t) &=& \sum_{m=0}^{\infty} \frac{(-1)^m\,4^m}{(2m+1)!}\,t^m, \quad\quad\quad
\Phi(t) =  \sum_{m=1}^{\infty}\frac{(-1)^{m+1}\,4^{m-1}}{(2\,m - 1)!\,m}\,t^m, 
\eea
in the definitions~\eqref{phirhodef}, we obtain the following expansions for $\rho_{ab}^{~~\,c}$ and $\pi_{ab}^{~~\,c}$:
\be
\rho_{ab}^{~~\,c} = \sum_{n=1}^{\infty}\rho_{~ab}^{(n)\,c},\quad\quad \pi_{ab}^{~~\,c} = \sum_{n=1}^{\infty}\pi_{~ab}^{(n)\,c}, \la{rhopiexp}
\ee
where the quantities 
\bea
\rho_{~ab}^{(n)\,c} = \left\{
\begin{array}{l}
\delta_b^c\, A_a \,\frac{(-1)^m\,4^{m-\frac{1}{2}}}{(2\,m+2)!}\, Z^m - (a\leftrightarrow b),\quad  n=2\,m+1,\quad m\in\mathbb{Z}_{+}, \\
\frac{1}{\lambda}\,\varepsilon_{ab}^{~~\,c}\,\frac{4^{m-2}\,(-1)^m}{(2\,m)!\,m}\,Z^m -  (a\leftrightarrow b),\quad  n=2\,m,\quad m\in\mathbb{N},
\end{array}
\right.
\eea
and
\be
\pi_{~ab}^{(n)\,c}  = \left\{
 \begin{array}{l}
  \delta_a^c\,A_b\,\frac{(-1)^m\,4^m}{(2\,m+1)!}\,Z^m - (a\leftrightarrow b),\quad n = 2\,m + 1, \quad m\in\mathbb{Z}_+, \\
  \big(-\lambda\,\varepsilon_{abp}\,A^pA^c + \frac{1}{\lambda}\varepsilon_{abc}\,Z\big)\,\frac{(-1)^{m-1}\,4^{m-1}}{(2m)!}\,Z^{m-1} - (a\leftrightarrow b) ,
  \quad n=2\,m, \quad m\in\mathbb{N}.
 \end{array}
 \right.
\ee
contain the $n$-th power of $A$.
Substituting the expansions~\eqref{rhopiexp} in the expression~\eqref{Fsu2}, we arrive at Eq.~\eqref{Fexp} with
\be
\mathcal{F}_{ab}^{(2m+1)} = 
\frac{(-1)^m\, 4^m}{(2m+1)!} \partial_a (A_b Z^m) + \frac{2}{\lambda}\,\frac{4^{m-2}\,(-1)^m}{(2\,m)!\,m}\,\varepsilon_{ab}^{~~\,c}\,\{Z^m,A_c\}
-(a\leftrightarrow b), \quad m \in\mathbb{Z}_+  \nonumber,
\ee
and
\bea
\mathcal{F}_{ab}^{(2m)} &=& \frac{(-1)^{m-1}\,4^{m-1} }{(2m)!} \left[ 
-\lambda \varepsilon_{abp}\partial_c(A^pA^c Z^{m-1}) + \frac{1}{\lambda}\varepsilon_{ab}^{~~\,c}\partial_c (Z^m)
 + \{A_a Z^{m-1},A_b\}\right] - (a\leftrightarrow b),\nonumber\\ 
 m&\in&\mathbb{N}.  \nonumber
\eea
Using the general rule~\eqref{linfF}, we get the following nontrivial $\ell$-brackets:
\bea
\ell_{2m+1}\left(A^{\otimes(2m+1)}\right) &=& 4^m \,\left[ \partial_a (A_b Z^m) + \frac{1}{\lambda}\,\frac{ (2m+1)}{8\,m}\,\varepsilon_{ab}^{~~\,c}\,\{Z^m,A_c\}\right]\,\dsf x^a\wedge \dsf x^b, \nonumber\\
\ell_{2m}\left(A^{\otimes(2m)}\right) &=& 4^{m-1} \left[ 
\lambda \varepsilon_{abp}\partial_c(A^pA^c Z^{m-1}) - \frac{1}{\lambda}\varepsilon_{ab}^{~~\,c}\partial_c (Z^m)
 - \{A_a Z^{m-1},A_b\}\right]\,\dsf x^a\wedge \dsf x^b , \nonumber\\
 m &\in& \mathbb{N},
\eea
which in addition to the brackets $\ell_2(f,g)$ and $\ell_{n+1}(f,A^{\otimes n})$ constructed in \cite[Section 6.1]{KS21} define the L$_\infty^{full}$ algebra corresponding to the $\mathfrak{su}(2)$-like Poisson gauge theory. The results of this section can be generalized to the four-dimensional case (with a commutative fourth coordinate) and/or to the angular non-commutativity \cite{DKKLV} using the explicit expressions for $\mathcal{F}$, presented in \cite{KurkovVitale22}.

\section{Comments on P$_\infty$}

The Poisson infinity or P$_\infty$ structures appeared in the context of quantization of co-isotropic sub-manifolds of Poisson manifolds \cite{Lyakhovich,CF}. In fact P$_\infty$ algebra is an L$_\infty$ algebra $(\ell_n)_{n=1}^\infty$ on $V$ in which, in addition, it is introduced an operation of the associative multiplication between the elements. To each pare of elements, $\forall u,v\in V$, it is assigned the third element $u\otimes v\to u\cdot v\in V$ in such a way that $(u\cdot v)\cdot w=u\cdot(v\cdot w)$. And the corresponding $\ell$-brackets should satisfy the Leibniz rule with respect to this product,
\begin{equation}\label{der}
\ell_n(v_1,\dots,v_{n-1},u\cdot v)=\ell_n(v_1,\dots,v_{n-1},u)\cdot v+u\cdot\ell_n(v_1,\dots,v_{n-1}, v)\,.
\end{equation}
Following \cite[Proposition 5.23]{KS21} we define the multiplication by truncating the
product on the exterior algebra at degree~$2$,
that is, 
\begin{align}\label{prod}
f\cdot g &= f\,g \ , \quad f\cdot A = f\, A\,, \qquad f\cdot E=f\,E\,,\\
  A\cdot B&=A\wedge B\,,\qquad A\cdot E=0\,,\qquad E\cdot F=0\,.\notag
\end{align}
for all zero-forms $f$ and $g$, one-forms $A$ and $B$ and two-forms $E$ and $F$. We remind that the spaces of higher forms are empty by construction. 

The same \cite[Proposition 5.23]{KS21} shows that the subalgebra of L$_\infty^{full}$ called L$_\infty^{gauge}$ which is concentrated in the degrees $0$ and $1$ and describes the action of the gauge symmetries on the gauge fields defines the P$_\infty$ algebra, i.e., the derivation properties of the corresponding brackets are satisfied,
\begin{eqnarray}
\ell_2(f,g\cdot h)&=&\ell_2(f,g)\cdot h+g\cdot\ell_2(f,h)\,,\\
 \ell_n\left(A^{\otimes n-1},f\cdot g\right)&=&\ell_n\left(A^{\otimes n-1},f\right)\cdot g+f\cdot\ell_n\left(A^{\otimes n-1},g\right)\,,\notag\\
 \ell_n\left(f,A^{\otimes n-2},g\cdot A\right)&=& \ell_n\left(f,A^{\otimes n-2},g\right)\cdot A+g\cdot  \ell_n\left(f,A^{\otimes n-1}\right)\,.\notag
\end{eqnarray}
To check whether L$_\infty^{full}$ algebra can be made into the P$_\infty$ structure we have to check the derivation properties (\ref{der}) on the $\ell$-brackets which belongs to the degree two, i.e., $\ell_2(f,E)$ and $\ell_n\left(A^{\otimes n}\right)$. The check involving $\ell_2(f,E)$ is straightforward. The Leibniz rule for the bracket $\ell_n\left(A^{\otimes n}\right)$ reads,
\begin{equation}\label{der1}
\ell_n\left(A^{\otimes n-1}, f\cdot A\right)=f\cdot\ell_n\left(A^{\otimes n}\right)+\ell_n\left(A^{\otimes n-1}, f\right)\cdot A\,,\qquad n>1\,.
\end{equation}
Using the explicit form of the bracket $\ell_n\left(A^{\otimes n}\right)$ given by (\ref{formulas}) the left hand side of (\ref{der1}) can be represented as,
\begin{eqnarray}\label{d1}
\ell_n\left(A^{\otimes n-1}, f\cdot A\right)&=&f\cdot\ell_n\left(A^{\otimes n}\right)+\\
&&\frac{n!}{2}\,(-1)^\frac{n(n-1)}{2}\left(P_{ab}{}^{cd(n-1)}\,A_d\,\partial_cf +R_{ab}{}^{cd(n-2)}\,A_d\left\{A_c,f\right\}\right)\dsf x^a\wedge \dsf x^b\,.\notag
\end{eqnarray}
According to the same formulas (\ref{formulas}) and (\ref{prod}) one finds,
\begin{eqnarray}\label{d2}
 \ell_2(A,f)\cdot A&=&-\left(\gamma^{c(1)}_a\,\partial_cf+\{A_a,f\}\right)A_b\,\dsf x^a\wedge \dsf x^b\,\\
\ell_n\left(A^{\otimes n-1}, f\right)\cdot A&=&(n-1)!\,(-1)^\frac{n(n-1)}{2}\,\gamma_a^{c(n-1)}\partial_cf\,A_b\,\dsf x^a\wedge \dsf x^b\,,\qquad n>2\,.\notag
\end{eqnarray}
In order (\ref{der1}) to hold the second term on the right of (\ref{d1}) should be equal to the right hand side of (\ref{d2}). For the constant NC parameter $\Theta^{ij}$, $\gamma_a^b(A)=\rho_a^b(A)=\delta_a^b$. In this case the non-vanishing brackets are $\ell_2(A,f)=-\{A_a,f\}\,\dsf x^a$ and $\ell_2(A,A)=-\{A_a,A_b\}\,\dsf x^a\wedge \dsf x^b$, while all higher brackets vanish. So, for the canonical non-commutativity the second term in the r.h.s. of (\ref{d1}) is precisely equal to (\ref{d2}) and the L$_\infty^{full}$ algebra turns into P$_\infty$ structure.

For non-constant $\Theta^{ij}(x)$ the situation is more complicated. The are non-vanishing higher brackets and the contribution to (\ref{d1}) containing the Poisson bracket $\{A_c,f\}$ will not be compensated by (\ref{d2}), since it does not have any term with the Poisson brackets. But the violation of (\ref{der1}) can be seen already  for $n=2$, since 
\begin{eqnarray}
P_{ab}{}^{cd(1)}\,A_d\,\partial_cf\,\dsf x^a\wedge \dsf x^b&=&-2\,\partial_a\Theta^{cd}A_d\,A_b\,\partial_cf\,\dsf x^a\wedge \dsf x^b\\
&\neq&-\frac12\,\partial_a\Theta^{cd}A_d\,A_b\,\partial_cf\,\dsf x^a\wedge \dsf x^b= \gamma^{c(1)}_a\,\partial_cf\,A_b\,\dsf x^a\wedge \dsf x^b\,.\notag
\end{eqnarray}
We conclude that the Leibniz rule in L$_\infty^{full}$ algebra holds true only in case of the canonical non-commutativity. Let us note also that L$_\infty^{gauge}$ algebra has has the structure of P$_\infty$ only for Poisson gauge transformations, for almost-Poisson case when the bi-vector $\Theta^{ij}(x)$ does not satisfy the Jacobi identity not even L$_\infty^{gauge}$ enjoys the derivation properties \cite{KS21}. At least so far we were not able to link the P$_\infty$ algebras to deformations of gauge theories. 

\section{Towards full non-commutative gauge theory}

As it has been already mentioned in the introduction the L$_\infty$-bootstrap approach is a powerful tool for the construction of order by order in $\Theta$ and $\hbar$ corrections of given undeformed gauge theory. Suppose we have a full non-commutative gauge theory with a corresponding L$_\infty$ algebra defined by the brackets $\ell^{NC}_n$ satisfying the relations (\ref{eq:calJndef}) schematically denoted by,
\begin{equation}\label{lir}
{\cal J}_n:= \sum^n_{i=1}\, (-1)^{i\,(n-i)} \ell^{NC}_{n-i+1}\ell^{NC}_i=0\,.
\end{equation}
According to the prescription of the L$_\infty$ bootstrap (\ref{i1}) and (\ref{i2}) the initial setup reads,
\begin{equation}\label{ic4}
\ell^{NC}_1(f)=\dsf f\,,\qquad\mbox{and}\qquad\ell^{NC}_2(f,g)=i[f,g]_\star\,.
\end{equation}
Let us set, $\ell_1(f)=\ell^{NC}_1(f)$, and
\begin{equation}\label{lim}
\ell_n=\lim_{\hbar\to0}\,\frac{1}{\hbar^{n-1}}\,\ell^{NC}_n\,,\qquad n>1\,.
\end{equation}
In particular,
\begin{equation}
\ell_2(f,g)=\lim_{\hbar\to0}\,\frac{1}{\hbar}\,[f,g]_\star=-\{f,g\}\,.
\end{equation}
If the brackets $\ell^{NC}_n$ satisfy the L$_\infty$ relations (\ref{lir}), then the brackets $\ell_n$ also do, and the brackets $\ell_1(f)$ and $\ell_2(f,g)$ are exactly the starting point for the L$_\infty$ algebra of Poisson gauge theory. In this sense we may call $\{\ell_n\}$ as a quasi-classical limit of $\{\ell^{NC}_n\}$ and the Poisson gauge theory is a quasi-classical limit of the full non-commutative gauge theory as it was mentioned in the introduction.
The brackets $\ell_n$ given by the Proposition \ref{pr3} provide the leading in $\hbar$ order of the brackets $\ell_n^{NC}$,
\begin{equation}
\ell^{NC}_n=\hbar^{n-1}\ell_n+{\cal O}(\hbar^n)\,.
\end{equation}
The question is how to construct the higher order in $\hbar$ contributions to the brackets $\ell^{NC}_n$. The aim of this section is to outline the solution to this problem within the L$_\infty$-bootstrap approach. 

For the future convenience we start introducing the condensed indices,
\begin{equation}
(i)^p=(i_1\dots i_p)\,,\qquad \left(\partial_i\right)^p=\partial_{i_1}\dots\partial_{i_p}\,,
\end{equation}
where the round bracket means symmetrization. In particular, 
\begin{equation*}
K^{(i)^pj}\left(\partial_i\right)^pA_j=K^{i_1\dots i_pj}\partial_{i_1}\dots\partial_{i_p}A_j\,.
\end{equation*}
The non-commutative deformation of the space-time is introduced through the given Hermitean, $(f\star g)^\ast=g^\ast\star f^\ast$, unital, $f\star 1=1\star f=f$, star product defined by the expression,
\begin{equation}
f\star g=f\,g+\sum_{n=1}^{\infty}\left(\frac{i\hbar}{2}\right)^nB_n(f,g)\,,\qquad B_1=\{f,g\}\,.
\end{equation}
The bi-differential operators \begin{equation}
B_n(f,g)=\sum_{k,l=1}^nB_n^{(i)^k|(j)^l}(\partial_{i})^kf\,(\partial_{j})^lg\,,
\end{equation} are determined to satisfy the associativity condition, $f\star(g\star h)=(f\star g)\star h$. The coefficient functions $B_n^{(i)^k|(j)^l}$ can be constructed following the prescription of the Formality Theorem \cite{Kontsevich}, or also using the poly-differential approach proposed in \cite{KV2008}. The Hermiticity of the star product implies that, $B_{2n}(f,g)=B_{2n}(g,f)$, and $B_{2n+1}(f,g)=-B_{2n+1}(g,f)$.
Thus the expression for the bracket $\ell_2^{NC}(f,g)$ becomes,
\begin{equation}
\ell^{NC}_2(f,g)=2i\sum_{n=0}^\infty \left(\frac{i\hbar}{2}\right)^{2n+1}B_{2n+1}(f,g)=-\hbar\,\{f,g\}+{\cal O}(\hbar^3)\,.
\end{equation}
To determine the bracket $\ell_2^{NC}(f,A)$ one has to solve the homotopy relation, ${\cal J}_2(f,g)=0$, written as,
\begin{equation}\label{e1}
\ell^{NC}_2(\ell^{NC}_1( f),g)+\ell_2^{NC}(f,\ell^{NC}_1( g))=\ell^{NC}_1(\ell_2^{NC}(f,g))\,.
\end{equation}
Remind that the element $\ell^{NC}_1(f)$ belongs to the space of the gauge fields, that is, we have a combination of yet undetermined brackets of the form $\ell_2(f,A)$ on the left expressed in terms of the combination of given brackets on the right. One may check that setting,
\begin{eqnarray}\label{l2fA}
\ell_2^{NC}(f,A)=\left\{i[f,A_a]_\star+\sum_{k,l=1}^{\infty}\tilde B_{a}^{(i)^k|(j)^l}(\partial_i)^kf\,(\partial_{j})^{l-1}A_{j_l}\right\}\dsf x_a\,,
\end{eqnarray}
where,
\begin{equation}
\tilde B_{a}^{(i)^k|(j)^l}=i\sum_{n=0}^\infty\left(\frac{i\hbar}{2}\right)^{2n+1}\partial_aB_{2n+1}^{(i)^k|(j)^l}\,,
\end{equation}
and defining according to the graded symmetry, $\ell_2^{NC}(A,f):=-\ell_2^{NC}(f,A)$, one solves (\ref{e1}).

The non-trivial technical issue appears on the following steps of the bootstrap procedure. Let us discuss the definition of the bracket $\ell_2^{NC}(A,B)$ entering the decomposition of the non-commutative field strength (\ref{Ffull}). It should be found from the identity, $ {\cal J}(f,A)=0$, written in the form,
\begin{eqnarray}
\label{g2}
\ell^{NC}_2(\ell^{NC}_1( f),A)+\ell^{NC}_2(f,\ell^{NC}_1( A))= \ell^{NC}_1( \ell^{NC}_2(f,A))\,.
\end{eqnarray}
The bracket, $\ell_2^ {NC}(f,A)$ on the right is given by (\ref{l2fA}), while on the left we have a combination of two yet undetermined brackets of the form $\ell_2^{NC}(A,B)$ and $\ell^{NC}_2(f,\ell^{NC}_1( A))$ reminding that the element $\ell^{NC}_1( A)$ belongs to the space space of gauge covariant objects $V_2$. Following the reasoning of the Proposition \ref{pr3} for the element ${\cal F}\in V_2$ we just set,
\begin{eqnarray}
\label{g4}
\ell^{NC}_2(f,{\cal F})=i[f,{\cal F}]_\star\,.
\end{eqnarray}
In this case, the associativity of the star product implies the Jacobi identity for the bracket (\ref{g4}), meaning that the homotopy relation, ${\cal J}_3(f,g,{\cal F})=0$, will be automatically satisfied and there is no need for the higher brackets $\ell_{n+2}^{NC}(f,{\cal F},A^n)$. The gauge covariance condition thus becomes, $\delta_f^{NC}{\cal F}=i[f,{\cal F}]_\star$.
Now we rewrite the eq. (\ref{g2}) as,
\begin{eqnarray}\label{e2}
&&\ell^{NC}_2(\ell^{NC}_1( f),A)={\cal J}_2^R(f,A)\,,\\
&&{\cal J}_2^R(f,A):=\ell^{NC}_1\left(\ell_2^{NC}(f,A)\right)-i[f,\ell^{NC}_1( A)]_\star\,.\label{e3}
\end{eqnarray}
Again, yet undetermined bracket on the left is expressed as a combination of the previously determined brackets on the right. One may schematically represent the r.h.s. as,
\begin{eqnarray}\label{e4}
{\cal J}_2^R(f,A)=\sum_{p,q=1}^\infty Q_{ab}{}^{(i)^p|(j)^{q-1}j_q}(\partial_{i})^p f\,(\partial_{j})^{q-1}A_{j_q}\,\dsf x^a\wedge \dsf x^b
\end{eqnarray}
where the coefficient functions $Q_{ab}{}^{(i)^p|(j)^{q-1}j_q}$ are constructed from the Poisson bi-vector $\Theta^{ij}(x)$ and its derivatives according to the prescription (\ref{e3}). The bracket $\ell_2^{NC}(A,B)$ can be also schematically represented as a sum,
\begin{equation}\label{e6}
\ell_2^{NC}(A,B)=\sum_{p,q=1}^\infty R_{ab}{}^{(i)^{p-1}i_p|(j)^{q-1}j_q}(\partial_{i})^{p-1} A_{i_p}\,(\partial_{j})^{q-1}B_{j_q}\,\dsf x^a\wedge \dsf x^b\,.
\end{equation}
The graded symmetry of the bracket, $\ell_2^{NC}(A,B)=\ell_2^{NC}(B,A)$, implies the symmetry relation of the corresponding coefficients,
\begin{equation}\label{e7}
R_{ab}{}^{(i)^{p-1}i_p|(j)^{q-1}j_q}=R_{ab}{}^{(j)^{q-1}j_q|(i)^{p-1}i_p}\,.
\end{equation}
To sum up, the coefficient functions $Q_{ab}{}^{(i)^p|(j)^{q-1}j_q}$ are given while $R_{ab}{}^{(i)^{p-1}i_p|(j)^{q-1}j_q}$ should be determined from the equation (\ref{e2}) in such a way that (\ref{e7}) holds true.

The bracket (\ref{e6}) will be constructed following \cite{Kup27}. First we observe that the graded symmetry of the bracket,
$
\ell_2^{NC}(\ell^{NC}_1( f),\ell^{NC}_1( g))=\ell_2^{NC}(\ell^{NC}_1( g),\ell^{NC}_1( f)),
$
implies the consistency condition on the right hand side of the equation (\ref{e2}),
\begin{eqnarray}
\label{cc3}
 {\cal J}_2^R(f,\ell^{NC}_1( g))-{\cal J}_2^R(g,\ell^{NC}_1( f))=\,0\,.
\end{eqnarray}
 The later however is automatically satisfied due to the previously solved homotopy relation, ${\cal J}_2(f,g)=0$, given by (\ref{e1}). Indeed, one calculates,
\begin{eqnarray}
\label{ce2}
\ell^{NC}_1\left(\ell_2^{NC}(f,\ell^{NC}_1( g))\right)- \ell^{NC}_1\left(\ell_2^{NC}(g,\ell^{NC}_1( f))\right)=
\ell^{NC}_1\left(\left[\ell^{NC}_1\left(\ell_2^{NC}(f,g)\right)-{\cal J}_2(f,g)\right]\right)\equiv\,0 \,.
\end{eqnarray}
The relation (\ref{cc3}) implies that,
\begin{eqnarray}
\sum_{p,q=1}^\infty Q_{ab}{}^{(i)^p|(j)^q} (\partial_i)^pf\,(\partial_j)^q g\,\dsf x^a\wedge \dsf x^b=\sum_{p,q=1}^\infty Q_{ab}{}^{(i)^p|(j)^q} (\partial_i)^pg\,(\partial_j)^q f\,\dsf x^a\wedge \dsf x^b\,,
\end{eqnarray}
which in turn results in the symmetry of the corresponding coefficient functions under the permutation of the symmetriezed groups of indices,
\begin{equation}\label{e5}
Q_{ab}{}^{(i)^p|(j)^q}=Q_{ab}{}^{(j)^q|(i)^p}\,.
\end{equation}
This relation is essential for the construction of the coefficients $R_{ab}{}^{(i)^{p-1}i_p|(j)^{q-1}j_q}$ with required symmetry properties, and it follows from the bootstrap procedure.
Note that the coefficient $Q_{ab}{}^{(i)^p|(j)^q}$ is symmetric over all $j$-indices, while the coefficient $Q_{ab}{}^{(i)^p|(j)^{q-1}j_q}$ is only symmetric with respect to first $(q-1)$ $j$-indices. So we introduce the notation for the ``not completely symmetric" part,
\begin{equation}\label{b10}
Q_{ab}{}^{(i)^p|[j]^q}:=Q_{ab}{}^{(i)^p|(j)^{q-1}j_q}-Q_{ab}{}^{(i)^p|(j)^q}\,,\qquad\mbox{with}\qquad Q_{ab}{}^{(i)^p|[j]^q}(\partial_j)^qf=0\,.
\end{equation}
It is also convenient to decompose $R_{ab}{}^{(i)^{p-1}i_p|(j)^{q-1}j_q}$ into maximally symmetric and partially antisymmetric parts,
\begin{eqnarray}\label{b14}
R_{ab}{}^{(i)^{p-1}i_p|(j)^{q-1}j_q}&=&R_{ab}{}^{(i)^{p}|(j)^{q}}+R_{ab}{}^{[i]^{p}|(j)^{q}}+\\
&&R_{ab}{}^{(i)^p|[j]^q}+R_{ab}{}^{[i]^{p}|[j]^{q}}.\notag
\end{eqnarray}
The symmetry relations (\ref{e7}) imply that,
\begin{eqnarray}\label{e8}
R_{ab}{}^{(i)^{p}|(j)^{q}}&=&R_{ab}{}^{(j)^{q}|(i)^p}\,,\\
R_{ab}{}^{[i]^{p}|(j)^{q}}&=&R_{ab}{}^{(j)^{q}|[i]^p}\,,\notag\\
R_{ab}{}^{[i]^{p}|[j]^{q}}&=&R_{ab}{}^{[j]^{q}|[i]^p}\,.\notag
\end{eqnarray}

Now consider the following contribution to the right hand side of the equation (\ref{e2}),
\begin{eqnarray}\label{e9}
\left[Q_{ab}{}^{(i)^p|(j)^{q-1}j_q}(\partial_{i})^p f\,(\partial_{j})^{q-1}A_{j_q}+Q_{ab}{}^{(j)^q|(i)^{p-1}i_p}(\partial_{j})^q f\,(\partial_{i})^{p-1}A_{i_p}\right]\dsf x^a\wedge \dsf x^b\,.
\end{eqnarray}
It should be compensated by the corresponding contribution to the bracket $\ell_2^{NC}(B,A)$ on the left,
\begin{eqnarray}\label{e10}
\left[R_{ab}{}^{(i)^{p-1}i_p|(j)^{q-1}j_q}(\partial_{i})^{p-1} B_{i_p}\,(\partial_{j})^{q-1}A_{j_q}+R_{ab}{}^{(j)^{q-1}j_q|(i)^{p-1}i_p}(\partial_{j})^{q-1}B_{j_q}(\partial_{i})^{p-1} A_{i_p}\right]\dsf x^a\wedge \dsf x^b,
\end{eqnarray}
with $ B_{i_p}=\partial_{i_p}f$. Substituting decompositions (\ref{b10}) and (\ref{b14}) in the eqs. (\ref{e9}) and (\ref{e10}) correspondingly and comparing the coefficients we conclude that,
\begin{eqnarray}
R_{ab}{}^{(i)^{p}|(j)^{q}}&=&Q_{ab}{}^{(i)^p|(j)^{q}}\,,\\
R_{ab}{}^{(i)^{p}|[j]^q}&=&Q_{ab}{}^{(i)^{p}|[j]^q}\,,\notag\\
R_{ab}{}^{[i]^{p}|(j)^{q}}&=&Q_{ab}{}^{(j)^{q}|[i]^p}\,,\notag
\end{eqnarray}
and since the coefficient $R_{ab}{}^{[i]^{p}|[j]^{q}}$ does not contribute to (\ref{e10}) we just set it to zero. We end up with,
\begin{equation}
R_{ab}{}^{(i)^{p-1}i_p|(j)^{q-1}j_q}=Q_{ab}{}^{(i)^p|(j)^{q}}+Q_{ab}{}^{(i)^{p}|[j]^q}+Q_{ab}{}^{(j)^{q}|[i]^p}\,.
\end{equation}
One may easily see that it satisfies the symmetry relation (\ref{e7}) because of (\ref{e5}).
And thus, we have proved the following,

{\proposition \label{pr4} The solution to the equation (\ref{e2}) satisfying the required symmetry property is given by,
\begin{eqnarray*}
\ell_2^{NC}(A,B)=\sum_{p,q=1}^\infty \left[Q_{ab}{}^{(i)^p|(j)^{q}}+Q_{ab}{}^{(i)^{p}|[j]^q}+Q_{ab}{}^{(j)^{q}|[i]^p}\right](\partial_{i})^{p-1} A_{i_p}\,(\partial_{j})^{q-1}B_{j_q}\,\dsf x^a\wedge \dsf x^b\,.\,\,\,\,\Box
\end{eqnarray*}}

To define the higher brackets one has to solve the higher homotopy relations given explicitly by the Proposition \ref{pr2}. Just like it happened in case of the Poisson gauge algebra the L$_\infty$ relation with three gauge parameters, ${\cal J}_{3}(f,g,h)=0$, is satisfied automatically because of the Jacobi identity for the bracket $\ell^{NC}_2(f,g)$ given by (\ref{ic4}). The latter means that there is no need in the higher brackets with two gauge parameters and we just set them to zero, $\ell^{NC}_{n+2}(f,g,A^{\otimes n})=0$. To determine the bracket with one gauge parameter and two gauge fields $\ell_3^{NC}(f,A,B)$ which enters the definition of the gauge variation $\delta_f^{NC}A$ one has to solve the L$_\infty$ relation, ${\cal J}_3(f,g,A)=0$. The latter can be written as,
\begin{eqnarray}\label{b6}
&&\ell_3^{NC}(\ell_1^{NC}(f),g,A)+\ell_3^{NC}(f,\ell_1^{NC}(g),A)={\cal J}^R_3(f,g,A)\,,\\
&&{\cal J}^R_3(f,g,A):=\label{b7}\\
&&-\ell_1^{NC}(\ell_3^{NC}(f,g,A))-\ell_2^{NC}(\ell_2^{NC}(f,g),A)-\ell_2^{NC}(\ell_2^{NC}(A,f),g)-\ell_2^{NC}(\ell_2^{NC}(g,A),f)\,,\notag
\end{eqnarray}
where yet undetermined brackets of the form $\ell_3^{NC}(f,A,B)$ on the left are expressed as a combination of the previously determined brackets ${\cal J}^R_3(f,g,A)$ on the right. The graded symmetry,
\begin{equation}
\ell_3^{NC}\left(f,\ell_1^{NC}(g),\ell_1^{NC}(h)\right)=\ell_3^{NC}\left(f,\ell_1^{NC}(h),\ell_1^{NC}(g)\right)\,,
\end{equation}
implies the consistency condition on the right hand side of (\ref{b6}),
\begin{equation}\label{cc4}
{\cal J}^R_3\left(f,g,\ell_1^{NC}(h)\right)+{\cal J}^R_3\left(g,h,\ell_1^{NC}(f)\right)+{\cal J}^R_3\left(h,f,\ell_1^{NC}(g)\right)=0\,.
\end{equation}
In \cite{Kupriyanov:2019ezf} it was demonstrated that this consistency condition holds true as a consequence of the previously solved homotopy relations, ${\cal J}_2(f,g)=0$, and ${\cal J}_3(f,g,h)=0$. An explicit form of the bracket  $\ell_3^{NC}(f,A,B)$ can be constructed following the logic of the Proposition \ref{pr4} and the relation (\ref{cc4}). The precise recurrence relations for the construction of full non-commutative (non-associative) gauge algebra were recently obtained in \cite{Kup-32}.

For some particular choices of the Poisson structure $\Theta^{ij}(x)$, like the $\mathfrak{su}(2)$-structure \cite{su(2)star1}-\cite{su(2)star5} or the $\kappa$-Minkowski \cite{kappastar1}-\cite{kappastar6} the expressions for the star product and the star commutator are known in all orders in $\hbar$. In this situation one may also look for the explicit expressions for the gauge transformations $\delta_f^{NC}$ and the corresponding dynamical objects ${\cal F}^{NC}$ and ${\cal D}^{NC}(\varphi)$. We leave it for the future investigation.

\section*{Acknowledgements}

The results of section 5 were obtained under support of the Tomsk State University Development Programe (Priority2030). O.A. thanks Universidade Federal do ABC (UFABC) and CAPES for support. V.G.K. was supported in parts  by the S\~ao Paulo Research Foundation (FAPESP), grant 2021/09313-8 and by the CNPq grant 304130/2021-4.

\end{document}